\documentclass[%
 aip,
 amsmath,amssymb,
 reprint,%
]{revtex4-1}

\usepackage{dcolumn}% Align table columns on decimal point
\usepackage{bm}% bold math
\usepackage[utf8]{inputenc}
\usepackage[T1]{fontenc}
\usepackage{mathptmx}
\usepackage{etoolbox}
\DeclareUnicodeCharacter{2212}{-} % Unicode U+2212 is MINUS SIGN, which is not set up by default with inputenc
\usepackage{subcaption}
\usepackage{float}%to fix the images at needed position we use [H] with beginfigure
\usepackage{graphicx}
\usepackage[final]{hyperref} % adds hyper links inside the generated pdf file

\hypersetup{
	colorlinks=true,       % false: boxed links; true: colored links
	linkcolor=blue,        % color of internal links
	citecolor=blue,        % color of links to bibliography
	filecolor=magenta,     % color of file linksa
	urlcolor=blue         
}

\usepackage[table,xcdraw]{xcolor} % to add text color
\usepackage{cleveref}
\usepackage{tikz}
\usepackage{xcolor} % For custom color definitions

\definecolor{jgreen}{HTML}{3ec800}
\definecolor{jorange}{HTML}{ec3500}
\definecolor{jbrown}{HTML}{cb9a00}
\definecolor{jblue}{HTML}{005e7f}
\definecolor{jpurple}{HTML}{cf1dff}

\usepackage{pdfpages}
\makeatletter\AtBeginDocument{\let\LS@rot\@undefined}
\makeatother

\usepackage{lineno}
\begin{document}

\preprint{AIP/123-QED}
% \linenumbers
\title{Recurrence analysis of meteorological data from climate zones in India}
% Force line breaks with \\

% \author[1]{Joshin John Bejoy}
% \author[2]{G.Ambika}
% \affil[1]{Indian Institute of Science Education and Research (IISER) Tirupati, Tirupati- 517507, India}

% \affil[2]{
% Indian Institute of Science Education and Research (IISERTVM), Thiruvananthapuram-695551 India%\\This line break forced% with \\
% }%

\author{Joshin John Bejoy}
\affiliation{Indian Institute of Science Education and Research (IISER) Tirupati, Tirupati- 517507 India}
\affiliation{Department of Aerospace Engineering, Indian Institute of Technology Madras, Chennai 600036, India}
%Lines break automatically or can be forced 
\author{G. Ambika*}
\affiliation{Indian Institute of Science Education and Research (IISERTVM), Thiruvananthapuram-695551 India
\\ *Corresponding Author: g.ambika@iisertvm.ac.in}

\date{\today}% It is always \today, today,

\begin{abstract}
We present a study on the spatiotemporal pattern underlying the climate dynamics in various locations spread over India, including the Himalayan region, coastal region, and central and northeastern parts of India. We try to capture the variations in the complexity of their dynamics derived from temperature and relative humidity data from 1948 to 2022. By estimating the recurrence-based measures from the reconstructed phase space dynamics using a sliding window analysis on the data sets, we study the climate variability in different spatial locations. The study brings out the variations in the complexity of the underlying dynamics as well as their heterogeneity across the locations in India. We find almost all locations indicate shifts to more irregular and stochastic dynamics for temperature data around 1972-79 and shifts back to more regular dynamics beyond 2000. These patterns correlate with reported shifts in the climate and Indian Summer Monsoon related to strong and moderate El Niño-Southern Oscillation events and confirm their associated regional variability.

\end{abstract}

\maketitle

\begin{quotation}
The climate system is known to be extremely complex, with the coexistence of many nonlinear interactions among its subsystems and several dynamical processes that change over spatial and temporal scales. This makes it difficult to model climate dynamics effectively. Hence, climate variability is studied by estimating measures of its complexity using nonlinear techniques on meteorological data like temperature, rainfall, and relative humidity. Among them, the recurrence-based approach in nonlinear time series analysis forms a very effective and powerful tool.
India is one of the countries that undergo significant changes in climate and exhibits highly heterogeneous climate variability. However, the climate system of India is not well studied for its spatial spread of these changes or its long-term temporal variations. In this study, we reconstruct the geometry underlying the climate dynamics from the temperature and relative humidity data of locations spread over the climate zones of India. We compute the measures from recurrence plots and recurrence networks using sliding window analysis over the data to understand the variations in climate dynamics in time in each location. This leads to the detection of shifts in climate variability and their variations across the locations. The study, thus presents the spatiotemporal variability in the dynamics related to major events in climate across the country. 
\end{quotation}

\section{\label{sec:level1}INTRODUCTION}

\begin{figure*}
    \captionsetup[subfigure]{justification=centering}
    \begin{subfigure}{2\columnwidth}
        \centering
        \includegraphics[width=\textwidth]{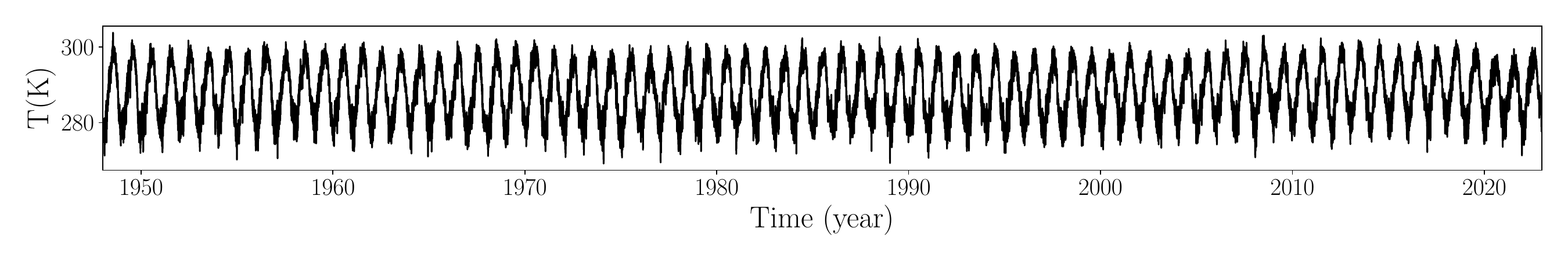}
        \caption{}
        \label{fig:2a}
    \end{subfigure}

    \begin{subfigure}{2\columnwidth}
        \centering
        \includegraphics[width=\textwidth]{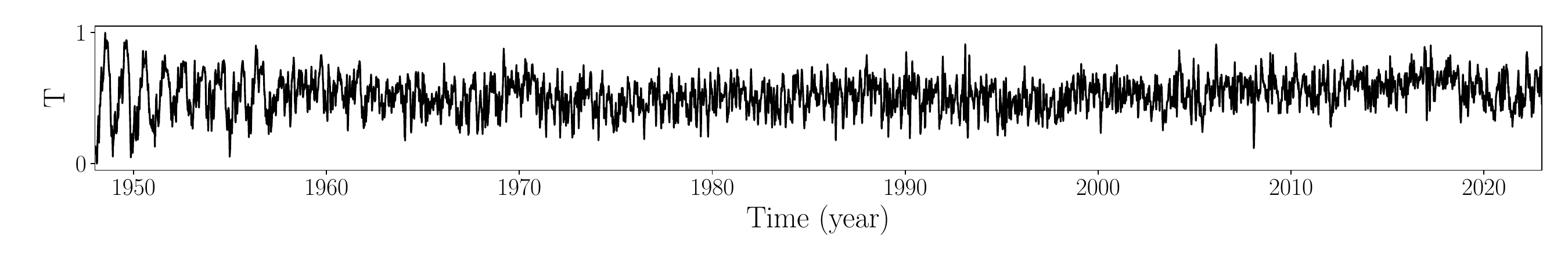}
        \caption{}
        \label{fig:2b}
    \end{subfigure}
    
    \begin{subfigure}{2\columnwidth}
        \centering
        \includegraphics[width=\textwidth]{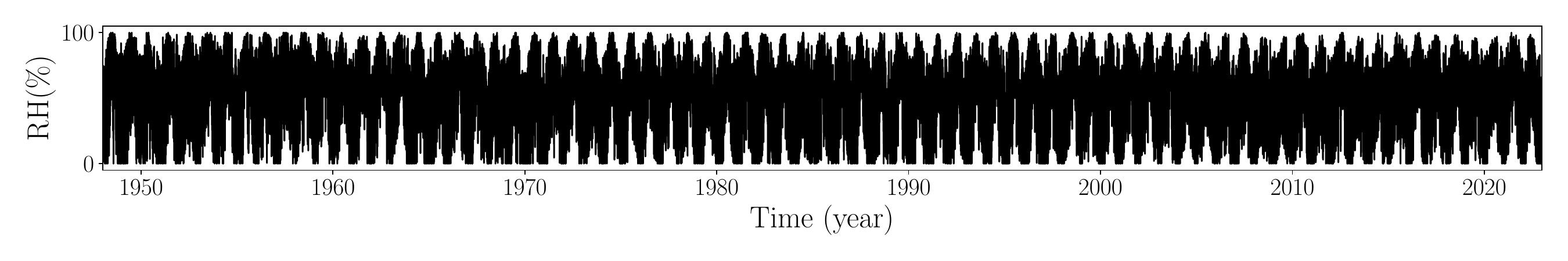}
        \caption{}
        \label{fig:2c}
    \end{subfigure}
    
    \begin{subfigure}{2\columnwidth}
        \centering
        \includegraphics[width=\textwidth]{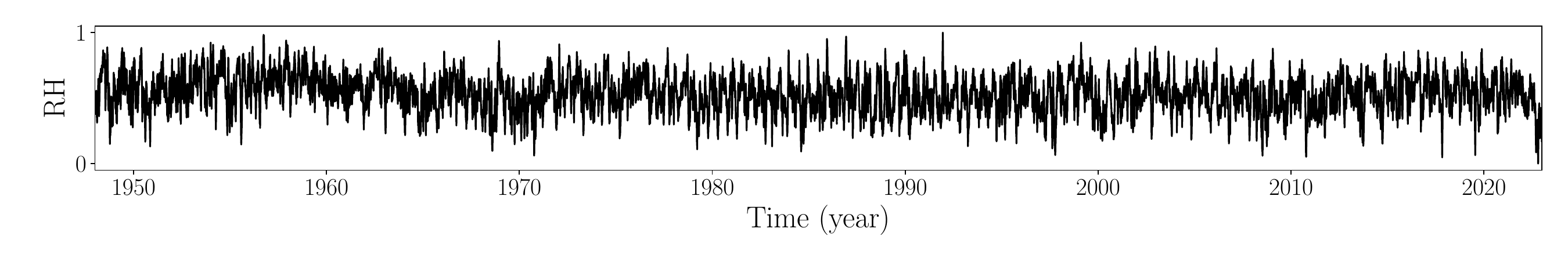}
        \caption{}
        \label{fig:2d}
    \end{subfigure}

    \caption{\textcolor{black}{Time series of temperature data from Manali: (a) original data and (b) data after preprocessing and rescaling. Time series of relative humidity data from Ladakh: (c) original data and (d) data after preprocessing and rescaling.}}
\label{fig:ts}
\end{figure*}

The climate is a highly nonlinear and heterogeneous dynamical system that exhibits complex variability over many scales in time and space. The variations in climate and global warming are of great concern as they affect humanity in many ways through reduction in agricultural yield, decline in water supplies, floods, erosion of coastal areas, droughts, changes in rainfall patterns, decrease in biodiversity, etc. Hence research related to climate and its variations are highly relevant for planning and policy-making for the benefit of humanity. We note that many interdisciplinary studies have been reported in this context in recent years \cite{1,2,3,4,5,6,7}. The major challenge in the study of climate variability arises due to the non-availability of dynamical equations describing the underlying processes. Hence, most of the studies rely on data of temperature, relative humidity, rainfall, percolation, etc., which require spatial details and temporal coverage. 

Several techniques are used in understanding complex systems, and among them, the method of nonlinear analysis is very effective in estimating complexity measures from observational or measured data\cite{8,9}. In this approach, the dynamics of the system is recreated from data, and the complexity of the dynamics is estimated using various techniques. Among them, the method of Recurrence Quantification Analysis (RQA) is well-accepted as a powerful tool for analyzing intricate patterns in data from various contexts\cite{10,11}. Compared to other methods, such as those based on Lyapunov exponents, this has the advantage that it can be effectively applied even with small and non-stationary data. Hence, this method is often adopted to study variations in complexity over time using a sliding window approach. The recurrence patterns in data are visualized as recurrence plots and represented as recurrence networks, and the quantifiers derived from them can provide information on transitions or changes in complexity over time\cite{12}. This has been successfully applied to study transitions in dynamics from astrophysical data, climate data, financial data, etc. \cite{13, 14, 15, 16, 17, 18, 19, 20, 21, 22}

India exhibits highly heterogeneous and complex variations in its climate. The high spatial heterogeneity arises basically from its geographical diversity but is influenced by the complex interactions between the atmosphere, ocean, land, and human interventions. The spatiotemporal variations of such effects can result in the manifestation of varied and complex dynamics in different locations in the country. However, an analysis of climate dynamics based on the average behavior at a national level will not bring out the variations that occur in different locations. Therefore, it is highly relevant to study the complexity of the spatiotemporal patterns of climate dynamics and identify locations that undergo changes during different climatic conditions.

In the context of the Indian climate, a few isolated studies are reported using rainfall and temperature data \cite{23, 24, 25, 26, 27}. However, a detailed study based on the nonlinear nature of climate variations over various locations in India is highly relevant. Due to the difference in geographical locations and also the difference in factors affecting their climates, like urbanization, industrialization, and population, we expect to have variations in the dynamics of their climate. 

We aim to identify the heterogeneity in the patterns and connections in the climate dynamics over the Indian subcontinent using data of temperature and relative humidity from locations across India over the period 1948 to 2022. Using the framework of nonlinear time series analysis, we recreate the dynamics underlying the climate system from these data. We do a sliding window analysis over the data in time and compute recurrence-based measures from recurrence plots and recurrence networks. From these, we try to detect shifts in climate in these locations based on the variability in their dynamics.\\
The present study is based on the structure of the underlying dynamics rather than on the statistical features of data and, thus, is novel to climate-related studies in India. Considering the high spatial heterogeneity and complexity of the Indian climate, the dynamical system’s approach appears relevant to reveal dynamical differences for a complete characterization of Indian climatology. As such, the study will have interesting insights and can contribute to our understanding of the changes in the dynamics of the climate system over the Indian subcontinent.

% Add rps and Phase spaces
\begin{figure*}[ht]
    \centering
    \subfloat[\centering]{{\includegraphics[width=\columnwidth]{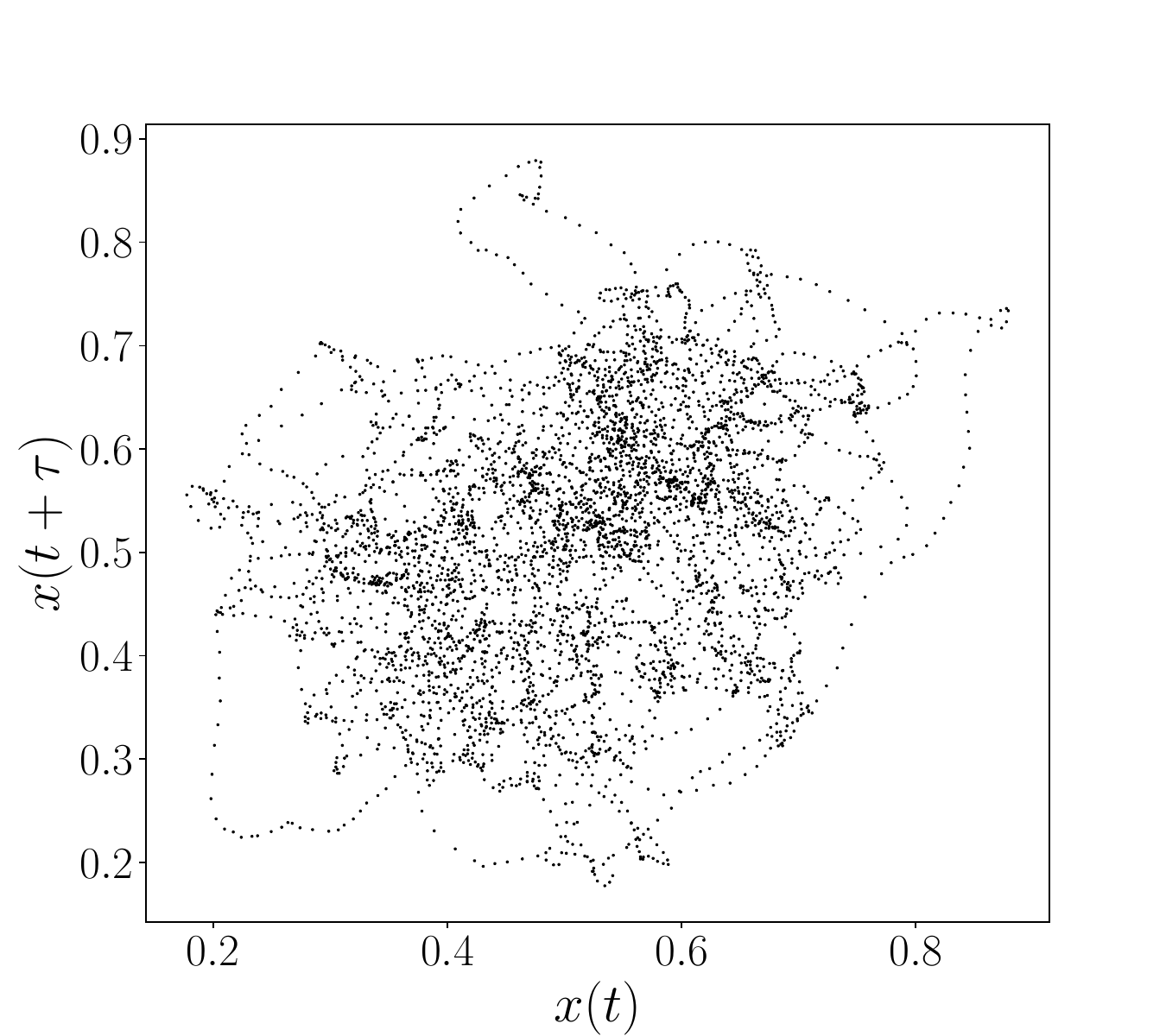} }}%
    % \caption{\label{fig:3a}}
    \label{fig:3a}
    \quad
    \subfloat[\centering]{{\includegraphics[width=\columnwidth]{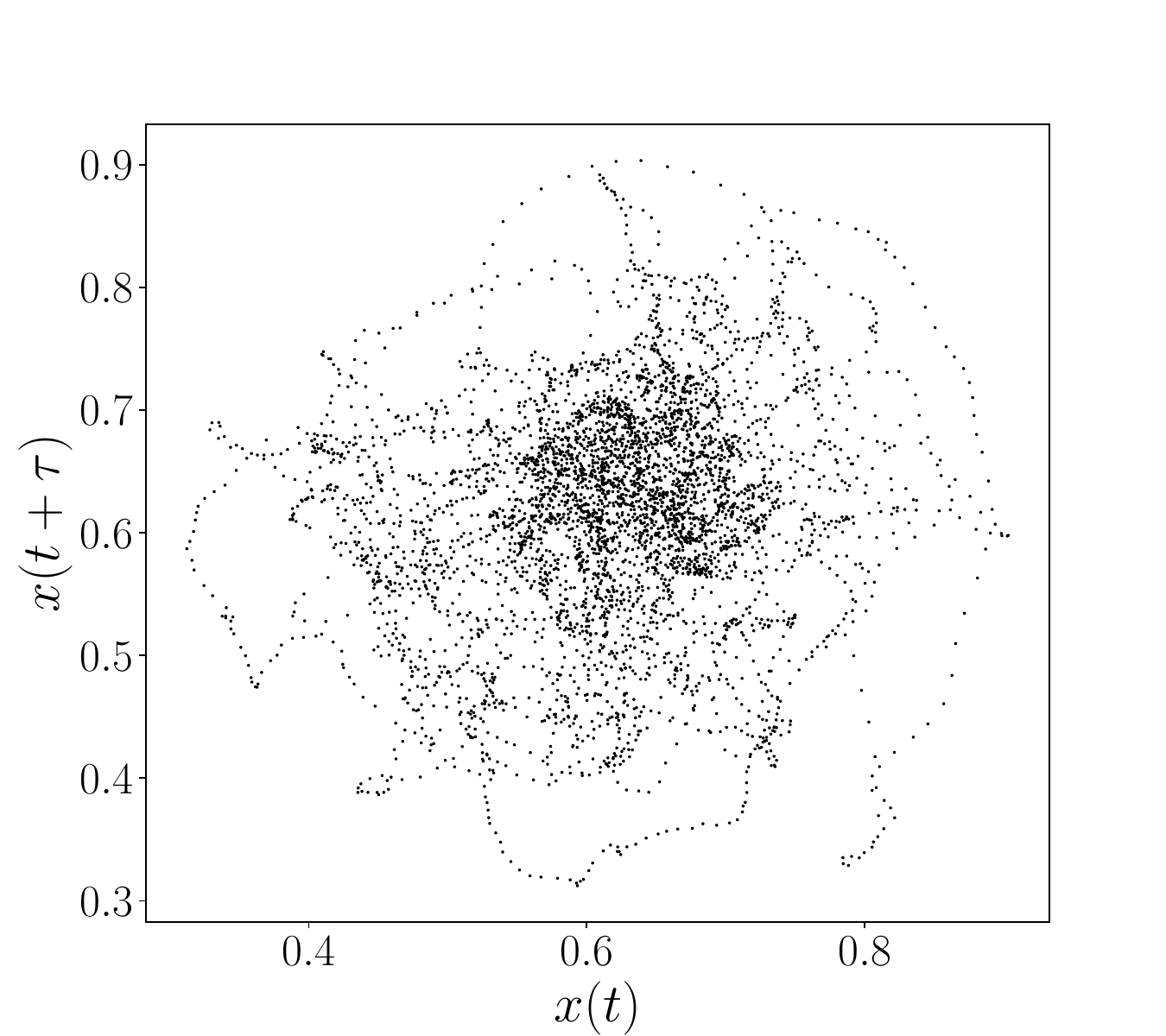} }}%
    % \caption{\label{fig:3b}}
    \label{fig:3b}

    % \subfloat[\centering]{{\includegraphics[width=\columnwidth]{3c TManaliwindow29 40.pdf}}}%
    % % \caption{\label{fig:3b}}
    % \label{3c}
    
    % \subfloat[\centering]{{\includegraphics[width=\columnwidth]{3d TManaliwindow94 40.pdf} }}%
    % % \caption{\label{fig:3b}}
    % \label{3d}

    \caption{\textcolor{black}{\textcolor{black}{2D projections of the reconstructed phase space trajectories obtained by the delay embedding of temperature data from Manali during (a) 1966-73, and (b) 2012-19.}}}
    \label{fig:ps}
\end{figure*}

\begin{figure*}
    \centering
    \subfloat[\centering]{{\includegraphics[width=\columnwidth]{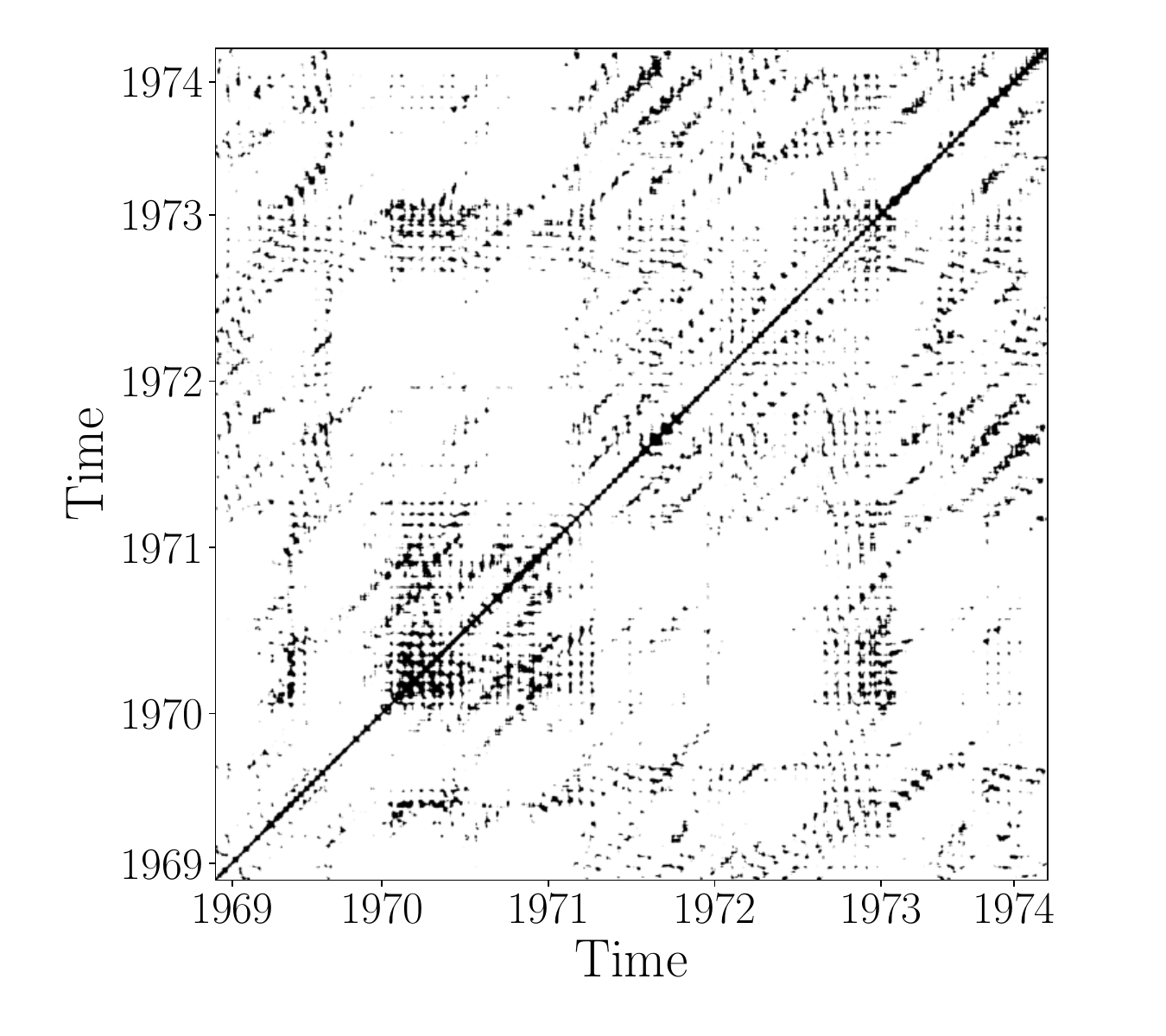} }}%
    % \caption{\label{fig:3a}}
    \label{fig:Mw1rp}
    \quad
    \subfloat[\centering]{{\includegraphics[width=\columnwidth]{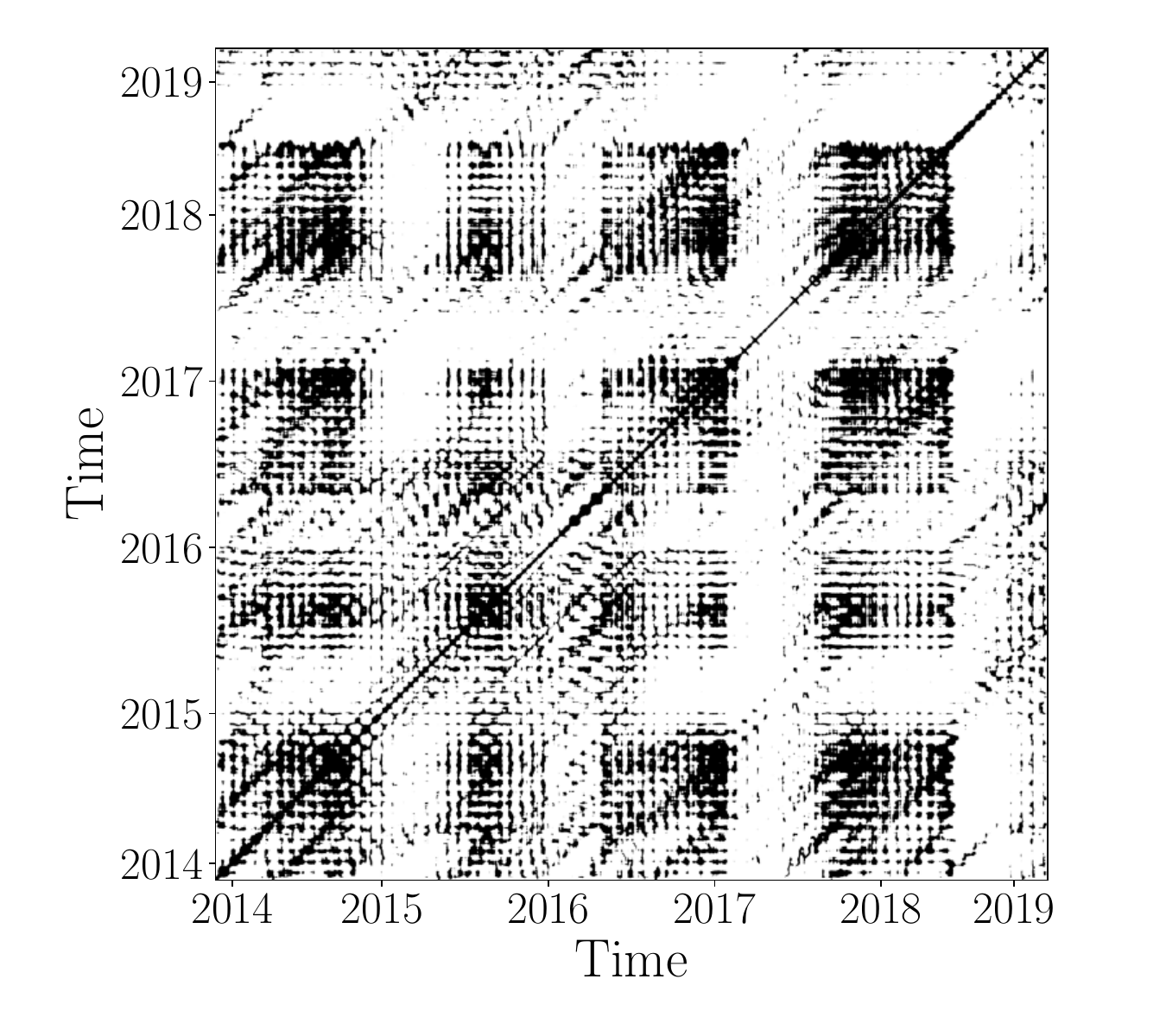} }}%
    % \caption{\label{fig:3b}}
    \label{fig:Mw2rp}
    \caption{\textcolor{black}{Recurrence plots from the phase space trajectories shown in FIG.\ref{fig:ps} for temperature data from Manali during the period (a) 1966-73 and (b) 2012-19.}}
    \label{fig:rp}
\end{figure*}

\section{Data and pre-processing}

We use reanalysis data sets of temperature and relative humidity from NCEP (National Centers for Environmental Prediction) gridded ($2.5^\circ\times 2.5^\circ$) (\url{https://psl.noaa.gov/}) for 15 locations across India over 75 years from 1948 to 2022. The 15 locations chosen for study from the different climate zones of India (based on the K\"{o}ppen-Geiger classification taken from Ref. \cite{28}) are Ladakh (La) Arid cold;
Shimla (Sh), Tawang (Ta) and Manali (Ma) Subtropical highland;
Patna (Pa), Mizoram (Mi), Ranchi (Ra), Bhopal (Bh) and Delhi (De) Humid sub-tropical;
Bengaluru (Be), Mumbai(Mu), Pondicherry (Po), and Tirupati (Ti) Tropical wet and dry;
Kannur (Ka) and Cochin (Co) Tropical wet.

As part of pre-processing, the original data are binned to get two points per day, and we remove the highest peak coming from annual periodicity. This is followed by filtering using moving-mean with a window of size of 30 to remove the seasonal components and random fluctuations. Finally, the data are rescaled to the range [0,1] before the analysis. The pre-processed and rescaled data of two typical locations, temperature from Ma and relative humidity from La, are shown in FIG.\ref{fig:ts}.

\section{Methodology}
As the first step in the analysis, we reconstruct the underlying dynamics from data using the method of delay embedding \cite{9}. For this, the scalar discrete time series $x(1), x(2), x(3),....,x(N)$ is embedded in an $M$-dimensional space by generating vectors with time delay coordinates, using a suitable time delay $\tau$ as
\begin{equation}
    \bar{X} = [x(i),x(i+\tau),...x(i+(M-1)\tau].
\end{equation}

The embedding is effective for appropriate choices of $M$ and $\tau$. Following the standard procedure, we take the time when autocorrelation $C(\tau)$ falls to $1/e$ as the appropriate delay time $\tau$. The minimum embedding dimension $M$ is estimated using the False Nearest Neighbours (FNNs) method\cite{8}, where false neighbors are identified using a chosen threshold for distances between points in the reconstructed trajectory in $M$ and $M+1$ dimensions. In our analysis, we take $M$=5, which is the maximum $M$ obtained for all the data sets. 

The recurrences of states in the reconstructed dynamics are captured as a two-dimensional image called recurrence Plot (RP), derived from the Recurrence matrix, $R$ defined as\cite{10}

\begin{equation}
    R_{i,j}=\Theta(\epsilon-||X_i-X_j||); i,j = 1,...,N,
\end{equation} where $N$ is the number of considered states $X_i$, $\epsilon$ is the threshold distance, $\lVert . \rVert$ is a norm, and $\Theta(.)$ is the Heaviside function.
The threshold $\epsilon$ is fixed as 0.2, which corresponds to the fifth percentile of the pairwise distances in the embedded five-dimensional space\cite{rcthreshold}. From the RPs constructed, we compute two standard measures: Determinism (DET), defined as the fraction of recurrence points that form diagonal lines, and Laminarity (LAM), which gives the fraction of recurrent points in vertical structures using the equations\cite{10}

\begin{equation}
    DET = \dfrac{\sum_{l=l_{min}}^NlP(l)}       
                {\sum_{l=1}^NlP(l)}
\end{equation}

\begin{equation}
    LAM = \dfrac{\sum_{v=v_{min}}^NvP(v)   }
                {\sum_{v=1}^NvP(v)}
\end{equation}
Here, $l$ is the length of diagonal lines, $v$ is the length of vertical lines, and $P(l)$ and $P(v)$ represent their frequency distributions. \textcolor{black}{By plotting $l\times P(l)$ vs $l$, we get the dominant length as the one corresponding to the first maximum of $l\times P(l)$. This gives the optimal minimal length to be included in the calculations. We suggest this as a method for choosing $l_{min}$, and it is found to work well for many data sets. Accordingly, we choose the minimum lengths $l_{min}$ and $ v_{min}$ as the value corresponding to the first maximum of $l\times P(l)$ and $v\times P(v)$, respectively.} In each case, the average value from all datasets is used in the calculations.

From the recurrence pattern of points on the reconstructed trajectory, the recurrence network (RN) is constructed by taking each point as a node and connecting two nodes by a link if the distance between them in the embedded space is
$\le \epsilon$. Then, the adjacency matrix $A$ of RN is obtained from $R$ by removing the diagonal elements (to avoid self-loops) as

\begin{equation}
    A_{ij}    =R_{ij}-\delta_{ij}
\end{equation}
The standard complex network measures, Link Density (LD) and Characteristic Path Length (CPL), are then computed within each window from the respective RNs using the following equations:
\begin{equation}
    LD =  \dfrac{2m}
                {N(N-1)}
\end{equation}where $m$ is the number of edges in the network and $N$ is the number of nodes,
% \begin{equation}
%     CPL = \sum _{\substack{s,t \in $v$ \\ s\ne t}}  \dfrac{d(s,t)}{n(n-1)}
% \end{equation}
\begin{equation}
    CPL = \dfrac{1}{N} \sum_{\substack{i}}^N \left(\dfrac{1}{N-1}\sum_{\substack{i\ne j \ne 1}}^{N-1}d_{ij}^s\right)
\end{equation}where $d_{ij}^s$ represents the shortest distance between nodes $i$ and $j$.

The recurrence-based measures characterize the recurrence pattern in the phase space trajectory. So, their changes can be interpreted in terms of the relative changes in climate dynamics. As established, high values of the measure DET correspond to regularity in the underlying dynamics, while low values indicate more irregular and stochastic variability in dynamics \cite{12,15}. The changes in the measure LAM can be interpreted as changes from one irregular state to another and can be correlated with changes in DET also in many cases.

The recurrence network measure LD increases when the dynamics become more regular and periodic but decreases for irregular and chaotic dynamics. On the other hand, the CPL values are small for regular dynamics and high for more complex and irregular dynamics \cite{20}. Thus, increasing trends in CPL values are an indication of a shift to more irregular dynamics. Also, an increase in LD corresponds to more regular dynamics that can change to an irregular nature as LD decreases.

The variations of these recurrence-based measures during the period of study can capture how the dynamics change over time in each location and the heterogeneity of such changes across the locations.

\begin{figure}[H]
    \centering
    \includegraphics[width=\columnwidth]{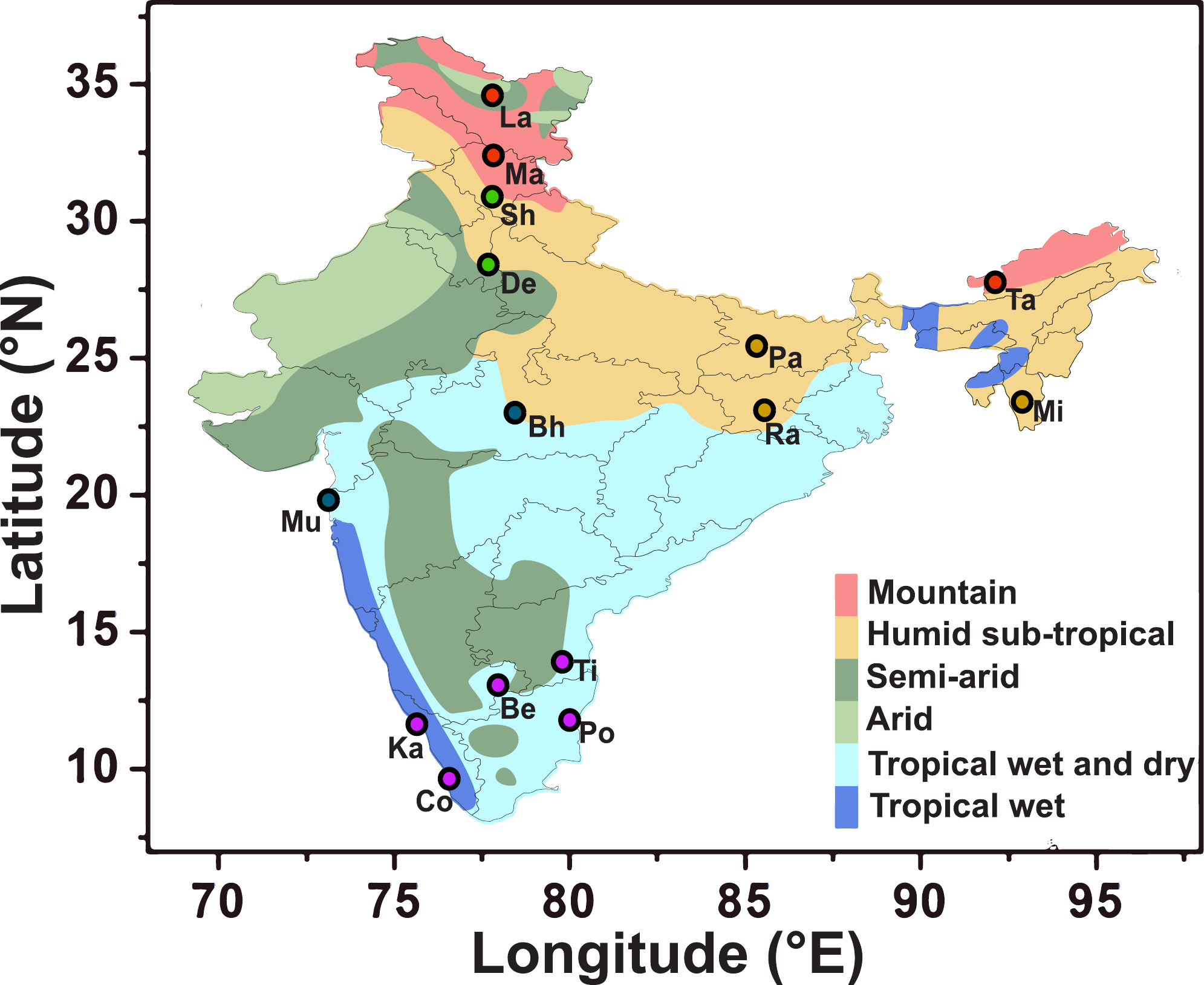}
    \caption{Locations grouped in five clusters based on the similarity in the variations of their DET measures from temperature data over the period of study. The clusters are differentiated by the colors used to locate them on the map of India. The major climate zones based on K\"{o}ppen-Geiger climate classification are also indicated\cite{28}.}
    \label{fig:tempclus}
\end{figure}

\begin{figure*}
    \captionsetup[subfigure]{justification=centering}
    \begin{subfigure}{2\columnwidth}
        \centering
        \includegraphics[width=0.9\textwidth]{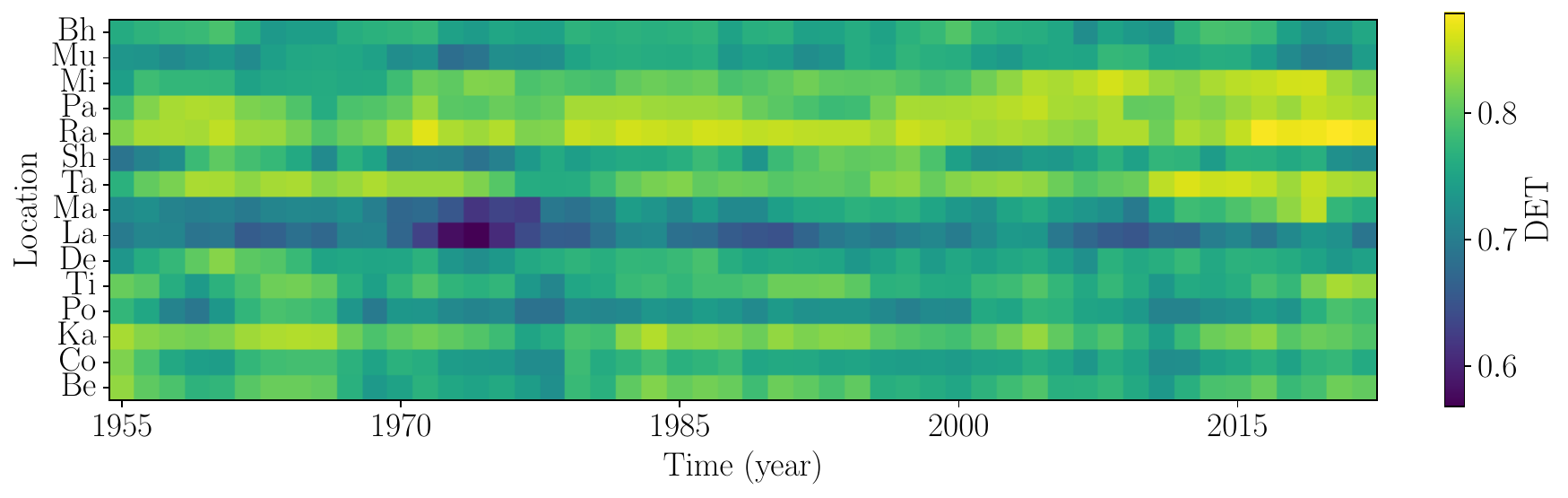}
        \caption{}
        \label{fig:dett}
    \end{subfigure}
    
    \begin{subfigure}{2\columnwidth}
        \centering
        \includegraphics[width=0.9\textwidth]{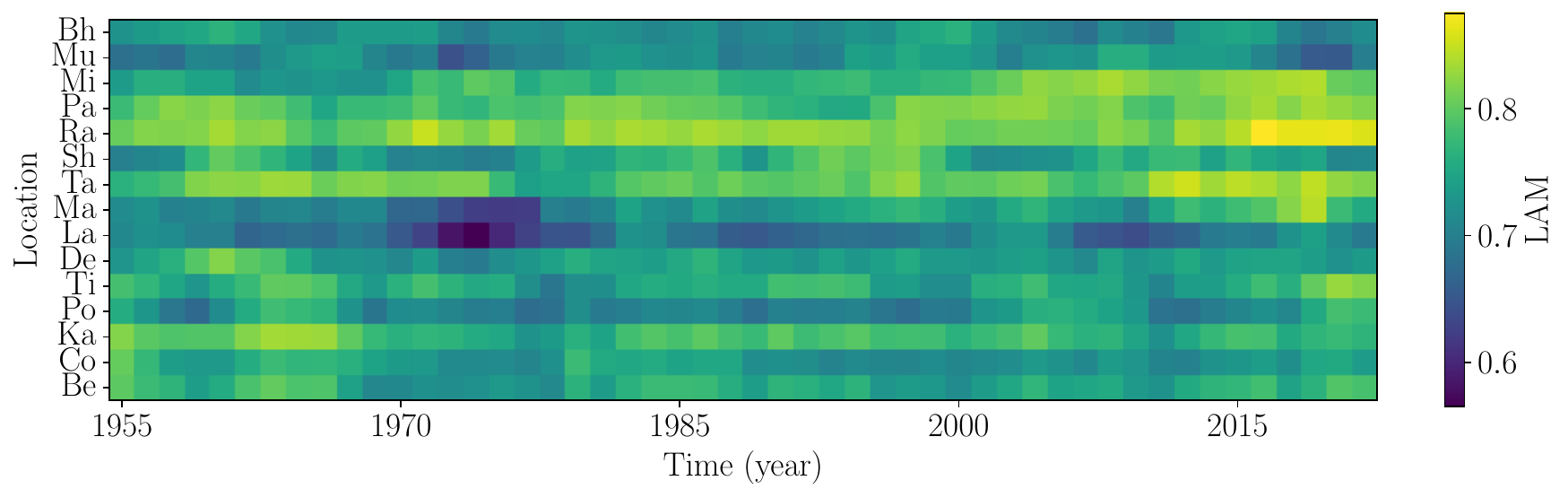}
        \caption{}
        \label{fig:lamt}
    \end{subfigure}
\caption{\textcolor{black}{Heatmaps of recurrence measures (a) DET and (b) LAM obtained from temperature data of the 15 locations. The locations having similar trends are grouped together.}}
\label{fig:rqat}
\end{figure*}

\section{Spatiotemporal variability in climate dynamics}
In addition to the inherent complexity in dynamics, climate is known to undergo significant changes or regime shifts with time\cite{29, 30, 31, 32, 33, 34}. This can result in different dynamical variability patterns in space and time. The reconstructed phase space trajectories for typical data of temperature from Ma for two different time periods are shown in FIG.\ref{fig:ps}. The corresponding recurrence plots are shown in FIG.\ref{fig:rp}. They clearly indicate changes in the underlying dynamics over time.

To understand such changes or regime shifts in the climate system, we do the sliding window analysis over the data for the period of study. The window considered in our study is 5000 time steps long ($\approx$7 years), and we slide it by 500 ($\approx$8 months) time steps. Then, we embed the data within each window and get the corresponding reconstructed attractor. We compute the recurrence measures DET and LAM as well as the recurrence network measures LD and CPL for the embedded attractor in each window. Their variations over time can then indicate the changes in the underlying dynamics and the corresponding transitions.

\textcolor{black}{We compute the Spearman cross-correlation matrix of the DET variations for the 15 locations and perform hierarchical agglomerative clustering\cite{spearman1,spearman2} using Euclidean metric, to get clusters based on similar variations in DET. Thus, for temperature data, the 15 locations fall into 5 clusters, with a silhouette score \cite{silhouette} of $0.28$.} The clusters, thus obtained are indicated in FIG.\ref{fig:tempclus} along with the climate zones. The locations in the same cluster can be identified by the colors used to locate them, as follows: 
Cluster 1(\tikz[baseline={([yshift=-0.65ex]current bounding box.center)}]\draw[black,fill=jpurple] (0,0) circle (0.5ex);): (Be, Co, Ka, Po, Ti), Cluster 2(\tikz[baseline={([yshift=-0.65ex]current bounding box.center)}]\draw[black,fill=jgreen] (0,0) circle (0.5ex);): (De, Sh), Cluster 3(\tikz[baseline={([yshift=-0.65ex]current bounding box.center)}]\draw[black,fill=jblue] (0,0) circle (0.5ex);): (Bh, Mu), Cluster 4(\tikz[baseline={([yshift=-0.65ex]current bounding box.center)}]\draw[black,fill=jorange] (0,0) circle (0.5ex);): (La, Ma, Ta) and Cluster 5(\tikz[baseline={([yshift=-0.65ex]current bounding box.center)}]\draw[black,fill=jbrown] (0,0) circle (0.5ex);): (Mi, Pa, Ra).

\textcolor{black}{The clusters do not follow the K\"{o}ppen-Geiger classification exactly. The locations in tropical wet and dry and tropical wet are in the same cluster based on their dynamical variations, except Mu, which is with Bh from the humid subtropical zone. Also, La, Ta, and Ma in arid cold and subtropical highland zones fall in the same cluster due to similar variations in dynamics. Pa, Ra, and Mi from the humid subtropical zone show similar dynamical variations, except Bh, while De and Sh form a separate cluster. The classification in this study is based on the underlying dynamics of each data that is affected by several climatic factors.}

\textcolor{black}{ We note that the values of LAM show similar trends in variations as DET in all locations. We present these variations as a heatmap, which comprehensively shows the values of each measure in the five clusters. The heatmaps, thus, derived from the measures of DET and LAM from temperature data for all locations are shown in FIGs. \ref{fig:dett} and \ref{fig:lamt}.}

\textcolor{black}{The clustering obtained is based on the variations in recurrence measures over the period of study. However, the significant increases or decreases in values of recurrence measures are to be located to identify regime shifts in climate dynamics. For this, following the methodology in \cite{tobias}, we perform a bootstrap resampling over the diagonal and vertical line length distributions of the recurrence plots and obtain a distribution of DET and LAM measures from all windows in each location. We select an inter-quantile range [0.1-0.9] of this distribution as the confidence bounds. The variations in values outside this range are considered significant changes in dynamics as captured by recurrence measures. The detailed plots with the variations in measures DET and LAM for each location are presented in the supplementary material.}

In FIG.\ref{fig:dett_scatt}, we indicate the durations when the significant variations in DET measures are seen, with decreasing trends shown in red and increasing trends in blue. \textcolor{black}{The vertical lines in red indicate the duration when most of the locations have decreasing trends in their DET values. The vertical line in blue beyond 1998 indicates the start of a period when DET values show increasing trends for most of the locations.} All locations except Pa, Ra, Mi, and Bh show decreases in DET during 1972-79, which correspond to variations in the underlying dynamics to a more irregular and stochastic nature. We note Pa, Ra, Mi, and Sh show a decrease earlier, and Bh shows variations in short intervals only after 2000. We find increasing trends in DET for most of the locations after 2000 that indicate a shift to more regular dynamics. However, Be, Co, Ka, and Po show increased DET values during 1955-66, Sh during 1993-97, and De during 1959-63.

\textcolor{black}{From the Oceanic Niño Index (ONI) data available online at \url{https://ggweather.com/enso/oni.htm}, we find within the two vertical lines in red, strong El Niño and La Niña alternate, with strong El Niño during 1972-73, followed by strong La Niña during 1973-76. The decreases in DET in most of the locations during 1972-79, which imply the shift in dynamics to a more irregular and stochastic nature, can correspond to occurrences of strong El Niño and La Niña events, while increases beyond 2000 to moderate and intermittent El Niño–Southern Oscillation (ENSO) events.}

\begin{figure}[H]
    \centering
    \includegraphics[width=\columnwidth]{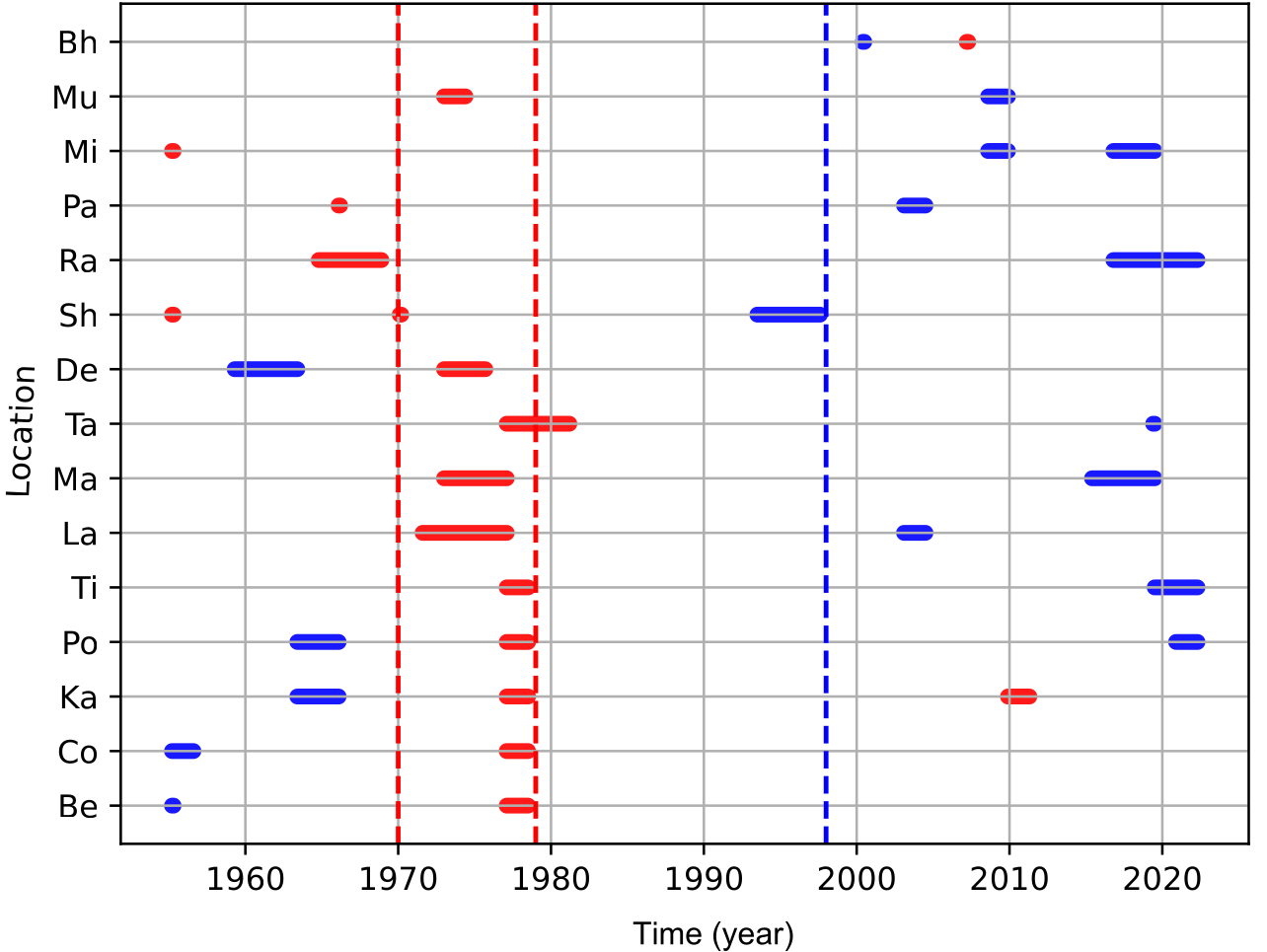}
    \caption{Durations when significant decreasing trends (red) and increasing trends (blue) are observed in DET measures from temperature data for the 15 locations. Within the two vertical lines in red most of the locations show decreasing trends in DET, while beyond the vertical line in blue DET values show increasing trends for most of the locations.}
    \label{fig:dett_scatt}
\end{figure}\noindent

\begin{figure}[H]
    \centering
    \includegraphics[width=\columnwidth]{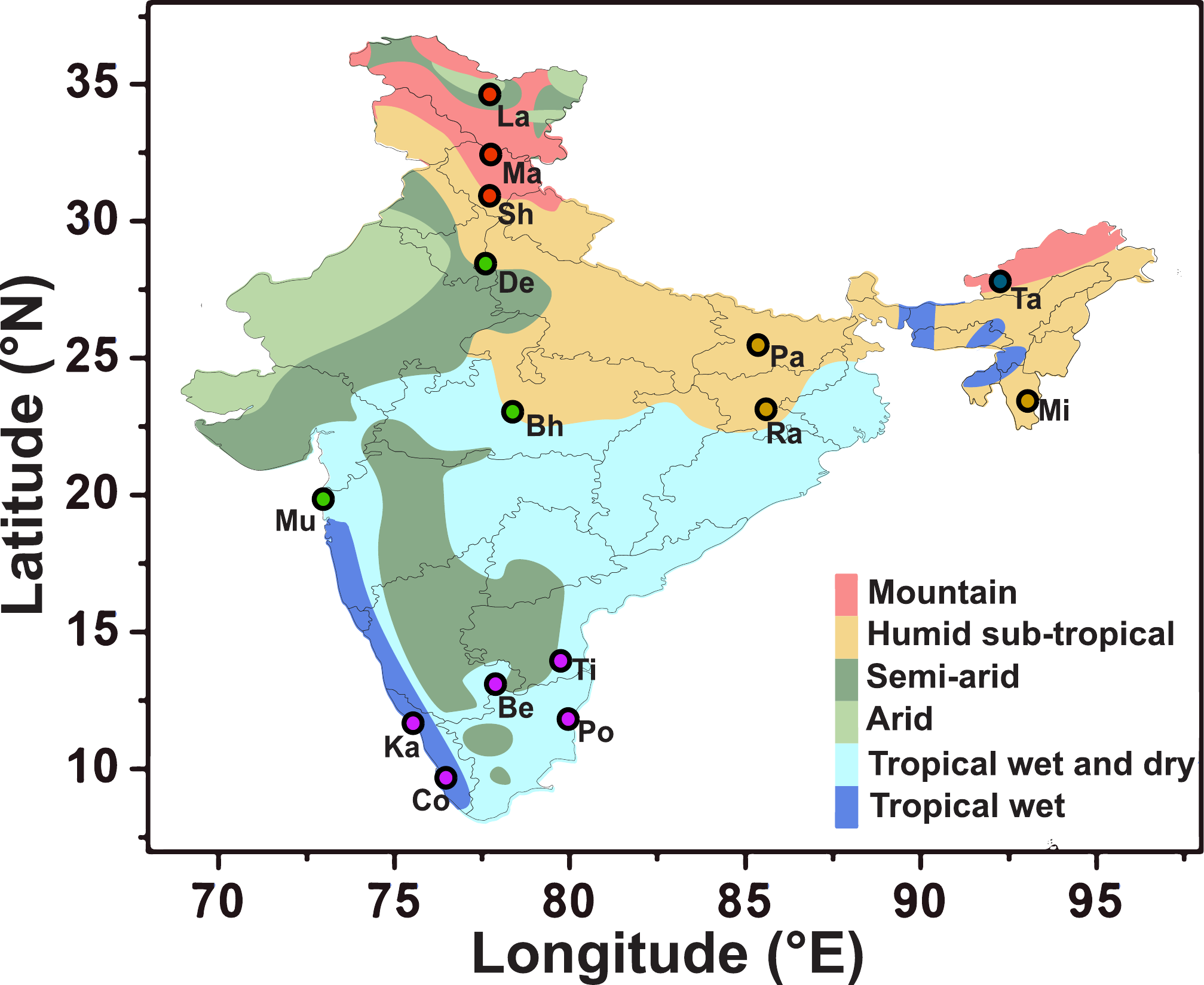}
    \caption{Locations grouped into five clusters based on the similarity in the variations of their DET measures from relative humidity data over the period of study. The clusters are indicated by the colors used to locate them on the map of India along with major climate zones.}
    \label{fig:humclus}
\end{figure}

\begin{figure*}
   \captionsetup[subfigure]{justification=centering}
    \begin{subfigure}{2\columnwidth}
        \centering
        \includegraphics[width=0.9\textwidth]{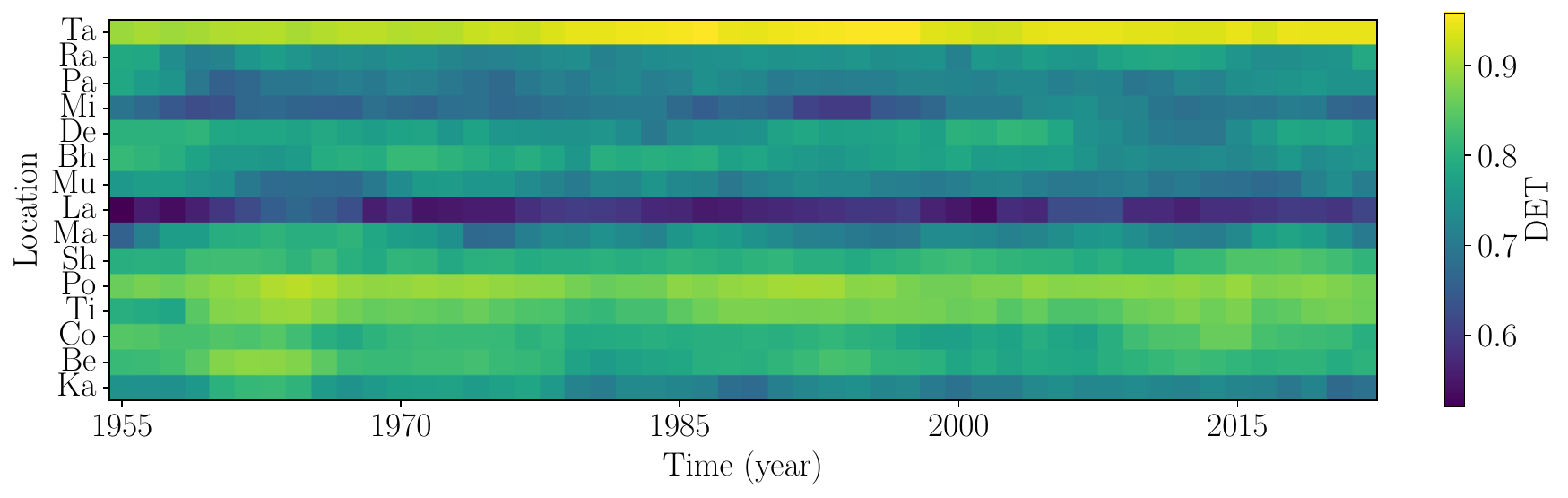}
        \caption{}
        \label{fig:deth}
    \end{subfigure}
    
    \begin{subfigure}{2\columnwidth}
        \centering
        \includegraphics[width=0.9\textwidth]{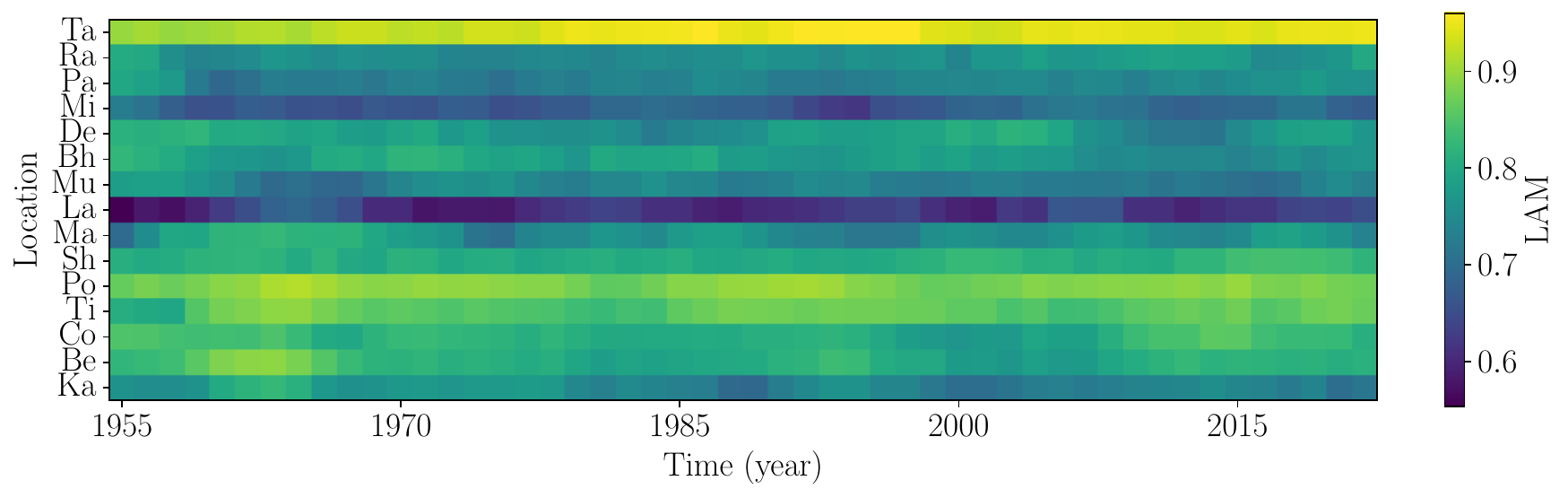}
        \caption{}
        \label{fig:lamh}
    \end{subfigure}
\caption{\textcolor{black}{Heatmaps of recurrence measures (a) DET and (b) LAM obtained from relative humidity data of the 15 locations. The locations having similar variations over time are grouped together.}}
    \label{fig:rqah}
\end{figure*}

% Add figs here scatterplot for DET -RH

\newpage
\begin{figure}[H]
    \centering
    \includegraphics[width=\columnwidth]{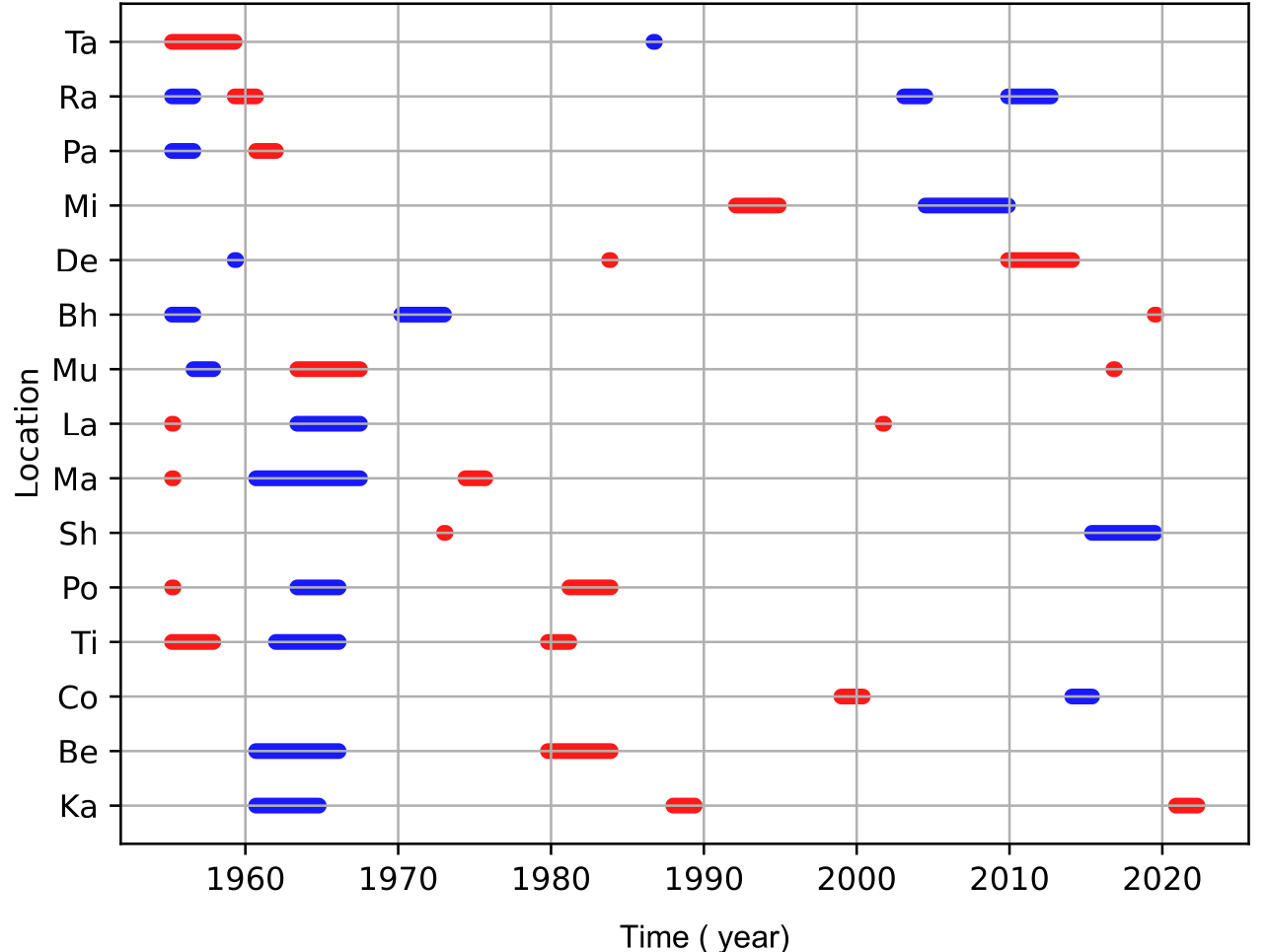}
    \caption{Durations when significant decreasing trends (red) and increasing trends (blue) are observed in DET measures from relative humidity data for the 15 locations. Before 1970 most of the locations show increasing trends in DET, while after that the decreasing trends seen are more heterogeneous and scattered with different durations. }
    \label{fig:deth_scatt}
\end{figure}

\textcolor{black}{Moreover, the spatial heterogeneity reported during such events can be related to variability in the duration and occurrence times of shifts in the underlying dynamics across the locations.}

\textcolor{black}{Based on the recurrence measures from data of relative humidity, we do similar clustering and get five clusters with a silhouette score of 0.19} as Cluster 1(\tikz[baseline={([yshift=-0.7ex]current bounding box.center)}]\draw[black,fill=jbrown] (0,0) circle (0.5ex);): (Mi, Pa, Ra), Cluster 2(\tikz[baseline={([yshift=-0.7ex]current bounding box.center)}]\draw[black,fill=jblue] (0,0) circle (0.5ex);): (Ta), Cluster 3(\tikz[baseline={([yshift=-0.7ex]current bounding box.center)}]\draw[black,fill=jgreen] (0,0) circle (0.5ex);): (Bh, De, Mu), Cluster 4(\tikz[baseline={([yshift=-0.7ex]current bounding box.center)}]\draw[black,fill=jorange] (0,0) circle (0.5ex);): (La, Ma, Sh), and Cluster 5(\tikz[baseline={([yshift=-0.7ex]current bounding box.center)}]\draw[black,fill=jpurple] (0,0) circle (0.5ex);): (Be, Co, Ka, Po, Ti). These are shown in FIG.\ref{fig:humclus}. \textcolor{black}{We find that the pattern of clustering is different in this case, with Ta forming a separate cluster, Bh, De, and Mu in the same cluster, La, Ma, and Sh forming another cluster, while Be, Co, Ka, Po, and Ti are together. We can identify the changes in dynamics in all locations from the heatmaps presented in FIGs.\ref{fig:deth}, \ref{fig:lamh} and for measures of DET and LAM from relative humidity data. The location-wise details of variations with confidence bounds are given in the supplementary material.}

\begin{figure*}[ht]
    \captionsetup[subfigure]{justification=centering}
    \begin{subfigure}{2\columnwidth}
        \centering
        \includegraphics[width=\textwidth]{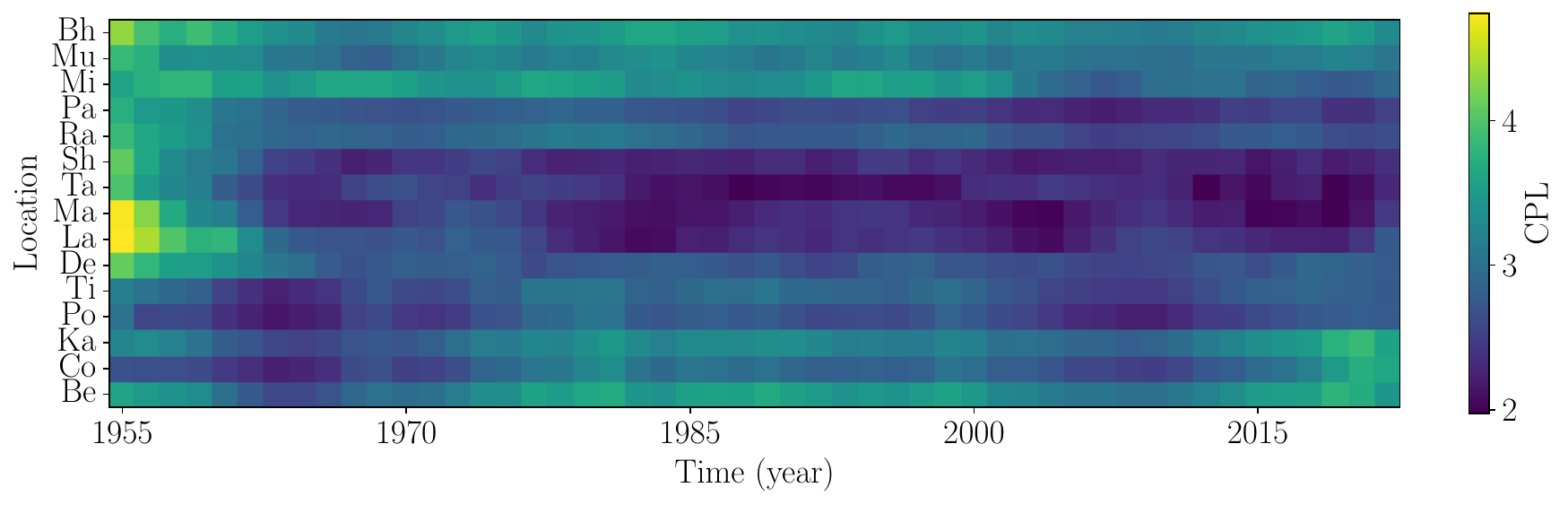}
        \caption{}
        \label{fig:cplt}
    \end{subfigure}
    
    \begin{subfigure}{2\columnwidth}
        \centering
        \includegraphics[width=\textwidth]{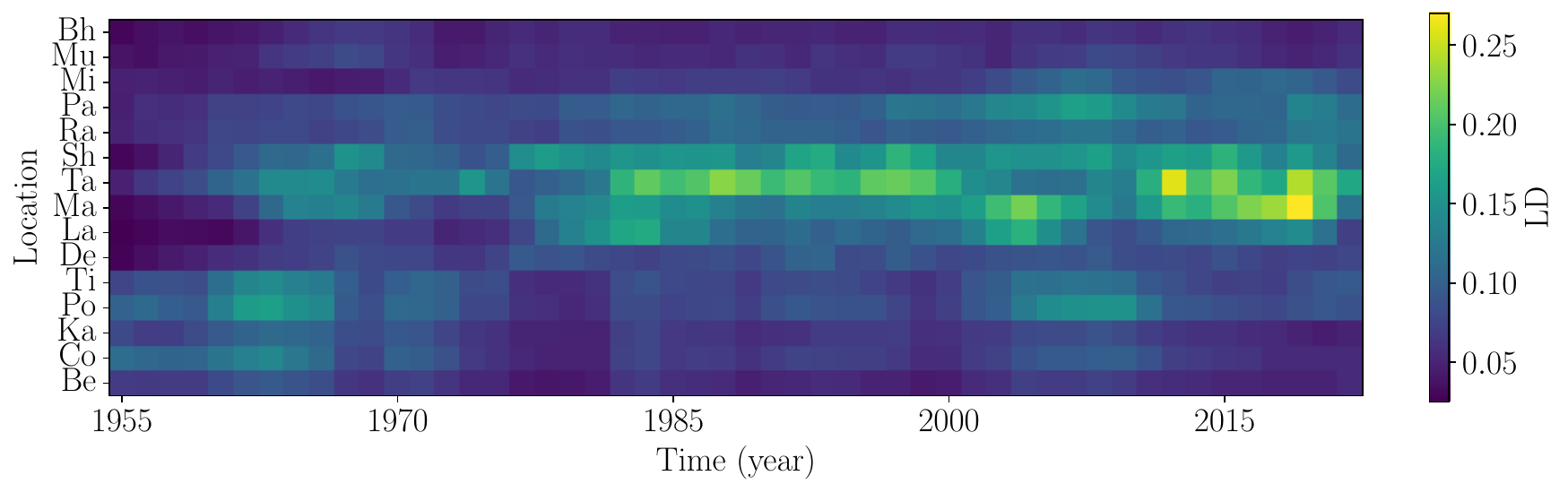}
        \caption{}
        \label{fig:ldt}
    \end{subfigure}

\caption{\textcolor{black}{Variations over time in recurrence network measures (a) CPL and (b) LD obtained from temperature data of the 15 locations displayed as heatmaps .}}
\label{fig:rnwt}
\end{figure*}

\begin{figure*}[ht]
    \captionsetup[subfigure]{justification=centering}
    \begin{subfigure}{2\columnwidth}
        \centering
        \includegraphics[width=\textwidth]{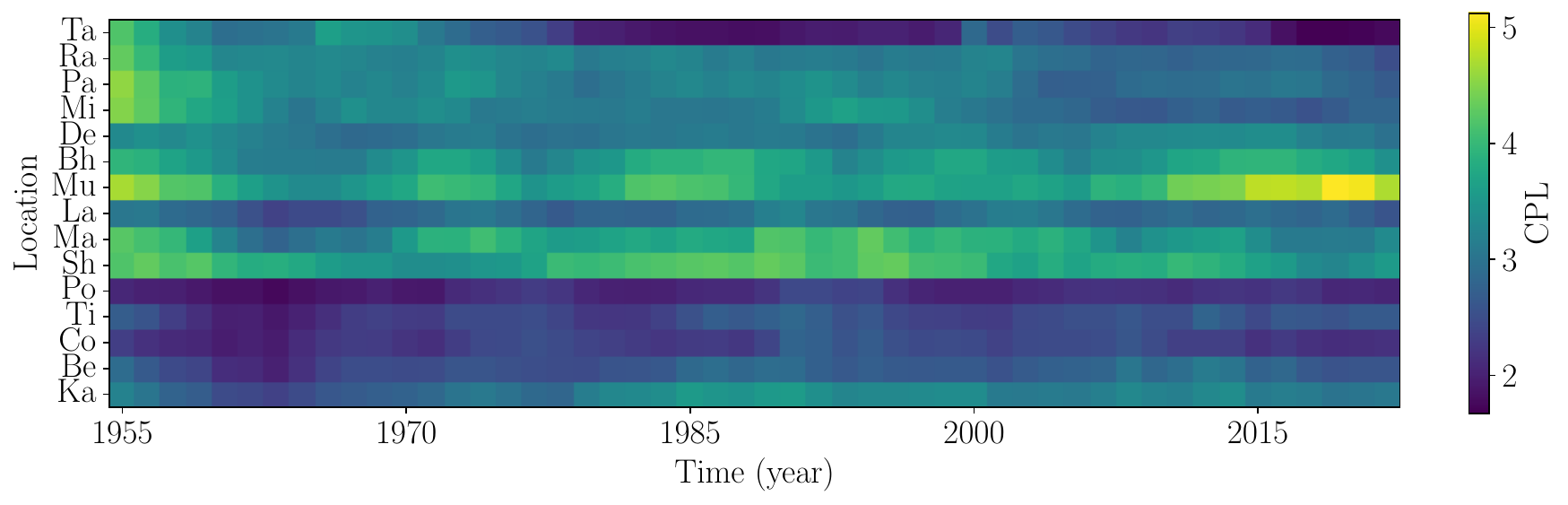}
        \caption{}
        \label{fig:cplh}
    \end{subfigure}
    
    \begin{subfigure}{2\columnwidth}
        \centering
        \includegraphics[width=\textwidth]{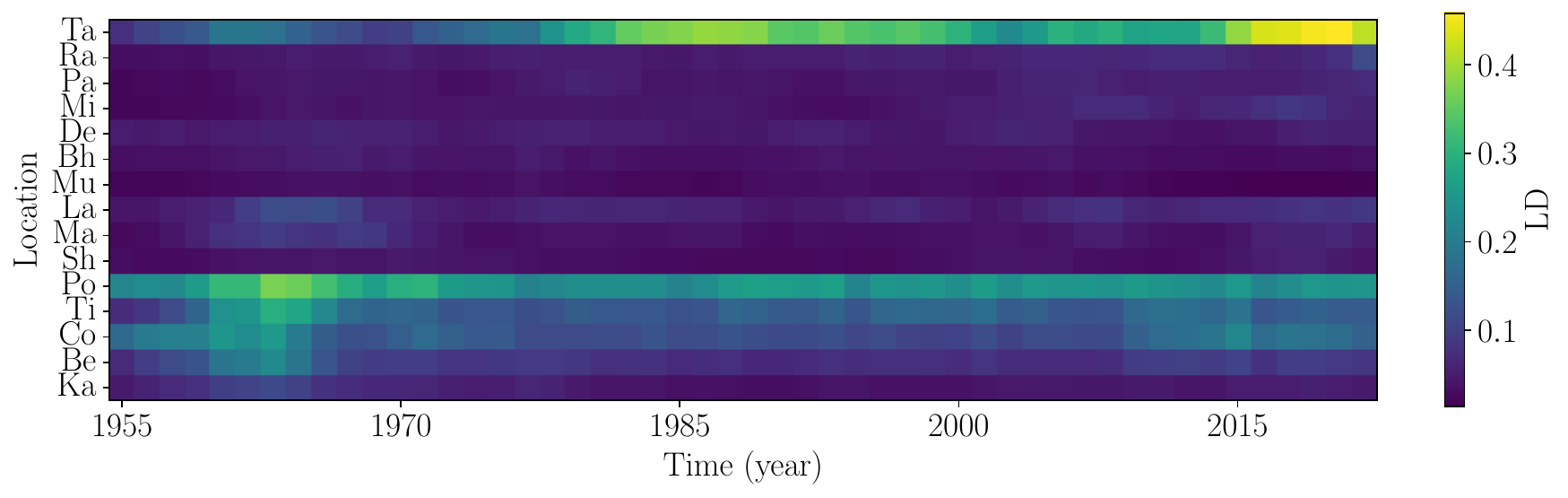}
        \caption{}
        \label{fig:ldh}
    \end{subfigure}

\caption{\textcolor{black}{Variations over time in recurrence network measures (a) CPL and (b) LD obtained from relative humidity data of the 15 locations displayed as heatmaps.}}
\label{fig:rnwh}
\end{figure*}

\textcolor{black}{For the dynamics underlying data of relative humidity, the significant variations in DET measures and their durations are indicated in FIG.\ref{fig:deth_scatt}. We find the variations are more heterogeneous and spread over longer durations. There are significant increases in DET for most of the locations except for Ta, Mi, Sh, and Co, before 1970, with decreasing trends after that at differing intervals across locations. Co shows a decrease in DET around 2000 and an increase around 2015. Ti has decreasing trends before 1960 and increases after that. We find increasing trends for Ra and Mi after 2000 and Sh after 2010. The variations at Ta are very different, with a decrease before 1960 and an increase around 1986.}

\textcolor{black}{The locations are grouped as per the variations in their CPL and LD values and are found to follow the same clustering as obtained from DET and LAM measures for both temperature and relative humidity data. These variations are shown as heatmaps in FIGs.\ref{fig:rnwt} and \ref{fig:rnwh} for temperature and relative humidity data, respectively. The details of the variations in these measures are presented in the supplementary material following the five clusters obtained from cross-correlation analysis of the variations over the period of study. They indicate cluster-wise variations and spatial variability across the locations.}

\textcolor{black}{We note most of the locations show an increase in CPL and a decrease in LD during 1970-79, indicating a shift to more irregular dynamics. Also, they show a decrease in CPL and an increase in LD after 2000, indicating reverse trends to more regular dynamics. However, the exact time of occurrence and duration of such changes vary location-wise, even within the same cluster, confirming the spatial variability across locations.}

% Add figs here scatterplot for DET -T

%%Add figs here heatmaps for DET and LAM-RH

% Add figs here heatmaps for CPL and LD-T
% Add figs here heatmaps for CPL and LD-RH

\section{Summary and conclusion}
India is known to have heterogeneity in climatic conditions over the different parts of the country. Therefore, it is of interest to study spatiotemporal variability to identify areas that show similar variations and those that differ in their responses to climatic signals.  
To understand the varying climate conditions and their complexity, the long-term meteorological time series from the different locations are to be analyzed and compared for their underlying dynamical nature and variations. In the present study, we report the analysis of temperature and relative humidity data from 15 different locations in India using the method of recurrence analysis. This study, based on a gridded dataset of $2.5^\circ \times 2.5^\circ$ resolution for temperature and relative humidity during the period 1948-2022, results in the detection of shifts in climate variability as inferred from recurrence patterns in the dynamics reconstructed from the data. The measures computed from recurrence plots and networks reveal the variations in the complexity over time in each location and the heterogeneity in these variations across the various locations.

The climate over the Indian subcontinent is influenced by nonlinear interactions between the atmosphere, ocean, and land that make the underlying dynamics highly complex. 
The temporal variations in the dynamics, as well as their spatial variations or heterogeneity over India, can be viewed as arising from the heterogeneous nature of several factors like spatial changes in atmospheric dynamics, wind circulations, and moisture and local factors, including human interventions, urbanization, industrialization, etc. \cite{40,41,42,43,44,45}

Based on the variations in recurrence measures that reflect variations in their dynamics over the period of study, we find the 15 locations can be grouped into 5 clusters. The clusters are not the same for temperature and relative humidity data, and spatial heterogeneity is evident in the dynamics reconstructed from both datasets. These clusters do not exactly follow the K\"{o}ppen-Geiger classification. The classification in this study is based on the recurrence pattern in the dynamics underlying the data rather than its statistical features or average values. 

We note that the dynamics will be affected by several climatic factors, such as the occurrences of ENSO episodes that have an impact on temperature patterns and wind circulation over the Indian subcontinent. The impact of these in precipitation and Indian Summer Monsoon (ISM) and their spatial variations over India during El Niño and La Niña periods are well studied \cite{39}. Their impact on climate dynamics is not uniform across India due to variations in atmospheric dynamics, wind circulations, etc.\cite{39} The ISM shows complex and time-varying interactions with the ENSO that can result in location-specific variations in temperature and relative humidity\cite{46}.

This study on the analysis of recurrence measures indicates statistically significant variations in the climate dynamics in general in all locations. We find that the changes observed are concurrent with reported major changes or regime shifts in India’s climate. Such changes emerge clearly from our analysis, with almost all locations showing shifts with a decrease in DET around 1972-79, with a few locations showing changes earlier from 1971-73. \textcolor{black}{It is possible that strong El Niño and La Niña that alternate during 1972-76, with their opposing influences on temperature, rainfall, pressure, etc. result in the climate dynamics becoming more irregular and stochastic. These results are also consistent with the climate shift reported during 1976-77 that has resulted in changes in ISM, as ENSO modulates Indian monsoon\cite{47}.}

Our study also indicates increases in DET measures beyond 2000 for most of the locations, indicating a shift to more regular dynamics. We note that, in the mid-1990s, an opposite climate shift was reported to the one that happened during 1976/77, and that also had an influence on the Indian monsoon onset\cite{48}. 
The shift in dynamics in most of the locations to more regular dynamics beyond 2000 \textcolor{black}{could be} influenced by the above factors. The changes in other measures, LAM, CPL and LD, also indicate similar changes in dynamics.

While the locations in each cluster show similar changes in their dynamics over the period of study, the times of occurrences and the significant variations differ based on the location, indicating spatial heterogeneity. In particular, Mu, De, and Bh show differences in their measures that seem to be influenced by local effects due to their geographical locations. 
In this context, we note that an increasing ENSO–ISM relationship is reported for north India, a decreasing relationship for central India after 1980, while it is a stable relationship for south India \cite{46}. \textcolor{black}{Thus, the heterogeneity in ISM and ENSO over different locations in India and the temporal variations in their relationship can cause spatiotemporal variability in their dynamics. However, there is no clear understanding on the spatial complexity of these trends and more involved future studies on climate dynamics are required in this context.}

The variations in recurrence measures derived from the dynamics of relative humidity data follow a clustering pattern different from that of temperature data, with Ta forming an isolated cluster and Mu, De, and Sh in the same cluster. The shifts in dynamics observed location-wise are also more heterogeneous and scattered with different durations, with regular dynamics in most of the locations before 1970 and irregular dynamics after that.
The dynamics underlying relative humidity seem to be affected by extra factors like precipitation, moisture transport, and sea surface temperature (SST) variations that lead to spatial heterogeneity in the seasonal variability\cite{49,50,51}. \textcolor{black}{Also, the differences between the spatial patterns in the dynamics of temperature and relative humidity data need to be studied further in detail and that opens up scope for future research.}

It is clear that the epochs of El Niño or La Niña, shifts in the pattern of ISM, and other climatological factors have an impact on the dynamics underlying temperature and relative humidity data, and they are not uniform across the locations in India. The present study indicates the importance of analyzing the dynamical variations underlying meteorological data from different grids to understand spatiotemporal heterogeneity rather than considering averages spread over the grids. The recurrence-based method applied in the study brings out the relevance of considering variations in the dynamics as it is capable of identifying variabilities in the dynamics over time. Thus, the transitions or shifts in climate dynamics can be understood and related to other major climatic events that occurred during the same period.

\section{ Supplementary material}
We provide details of variations in recurrence measures in each of the locations in the supplementary material. The figures showing variations in DET and LAM values over time in each location are given with details. Also, we present variations in the measures of CPL and LD derived from their recurrence networks, grouped as five sets of locations.

% \begin{figure*}
%     \centering
%     \includegraphics{1 T og spearman unified fsize 12 LAM 170x237mm new xlabels.pdf}
%     \caption{Variations in the recurrence measures DET and LAM in each of the 15 locations from
% temperature data. The selected quantiles in the range (0.01,0.99) are indicated as the
% shaded region, and the variations outside that are considered significant.}
%     \label{fig:alldett}
% \end{figure*}

% \begin{figure*}
%     \centering
%     \includegraphics{2 H og spearman unified fsize 12 LAM 170x237mm new xlabels.pdf}
%     \caption{Variations in the recurrence measures DET and LAM in each of the 15 locations from
% relative humidity data.}
%     \label{fig:alldeth}
% \end{figure*}

% \begin{figure}
%     \centering
%     \includegraphics{}
%     \caption{Caption}
%     \label{fig:enter-label}
% \end{figure}

\section{Data and source codes}
The data used in the study are the reanalysis data sets from NCEP (National Centers for Environmental Prediction) downloaded from \url{https://psl.noaa.gov/}.
For the computation of measures from the Recurrence Plot, we used the software available at \url{https://github.com/pik-copan/pyunicorn} \cite{pyuni}.

\section{Acknowledgments}
The authors thank Dr. Sai Kranthi, IISER Tirupati, and Dr. Bedartha Goswami, University of T\"{u}bingen, for helpful discussions. JJB acknowledges IISER Tirupati for facilities during the project work.

% \nocite{*}
% \bibliographystyle{plain}
\renewcommand{\refname}{References}
\bibliography{main}% Produces the bibliography via BibTeX.

\providecommand{\noopsort}[1]{}\providecommand{\singleletter}[1]{#1}%
\begin{thebibliography}{53}%
\makeatletter
\providecommand \@ifxundefined [1]{%
 \@ifx{#1\undefined}
}%
\providecommand \@ifnum [1]{%
 \ifnum #1\expandafter \@firstoftwo
 \else \expandafter \@secondoftwo
 \fi
}%
\providecommand \@ifx [1]{%
 \ifx #1\expandafter \@firstoftwo
 \else \expandafter \@secondoftwo
 \fi
}%
\providecommand \natexlab [1]{#1}%
\providecommand \enquote  [1]{``#1''}%
\providecommand \bibnamefont  [1]{#1}%
\providecommand \bibfnamefont [1]{#1}%
\providecommand \citenamefont [1]{#1}%
\providecommand \href@noop [0]{\@secondoftwo}%
\providecommand \href [0]{\begingroup \@sanitize@url \@href}%
\providecommand \@href[1]{\@@startlink{#1}\@@href}%
\providecommand \@@href[1]{\endgroup#1\@@endlink}%
\providecommand \@sanitize@url [0]{\catcode `\\12\catcode `\$12\catcode `\&12\catcode `\#12\catcode `\^12\catcode `\_12\catcode `\%12\relax}%
\providecommand \@@startlink[1]{}%
\providecommand \@@endlink[0]{}%
\providecommand \url  [0]{\begingroup\@sanitize@url \@url }%
\providecommand \@url [1]{\endgroup\@href {#1}{\urlprefix }}%
\providecommand \urlprefix  [0]{URL }%
\providecommand \Eprint [0]{\href }%
\providecommand \doibase [0]{http://dx.doi.org/}%
\providecommand \selectlanguage [0]{\@gobble}%
\providecommand \bibinfo  [0]{\@secondoftwo}%
\providecommand \bibfield  [0]{\@secondoftwo}%
\providecommand \translation [1]{[#1]}%
\providecommand \BibitemOpen [0]{}%
\providecommand \bibitemStop [0]{}%
\providecommand \bibitemNoStop [0]{.\EOS\space}%
\providecommand \EOS [0]{\spacefactor3000\relax}%
\providecommand \BibitemShut  [1]{\csname bibitem#1\endcsname}%
\let\auto@bib@innerbib\@empty
%</preamble>
\bibitem [{\citenamefont {Ghil}\ and\ \citenamefont {Lucarini}(2020)}]{1}%
  \BibitemOpen
  \bibfield  {author} {\bibinfo {author} {\bibfnamefont {M.}~\bibnamefont {Ghil}}\ and\ \bibinfo {author} {\bibfnamefont {V.}~\bibnamefont {Lucarini}},\ }\bibfield  {title} {\enquote {\bibinfo {title} {The physics of climate variability and climate change},}\ }\href {http://dx.doi.org/doi:10.1103/RevModPhys.92.035002} {\bibfield  {journal} {\bibinfo  {journal} {Rev. Mod. Phys}\ }\textbf {\bibinfo {volume} {92}},\ \bibinfo {pages} {035002} (\bibinfo {year} {2020})}\BibitemShut {NoStop}%
\bibitem [{\citenamefont {Dash}\ and\ \citenamefont {Mamgain}(2011)}]{2}%
  \BibitemOpen
  \bibfield  {author} {\bibinfo {author} {\bibfnamefont {S.~K.}\ \bibnamefont {Dash}}\ and\ \bibinfo {author} {\bibfnamefont {A.}~\bibnamefont {Mamgain}},\ }\bibfield  {title} {\enquote {\bibinfo {title} {Changes in the frequency of different categories of temperature},}\ }\href {http://dx.doi.org/doi:10.1175/2011JAMC2687.1} {\bibfield  {journal} {\bibinfo  {journal} {J. Appl. Meteorol. Climatol.}\ }\textbf {\bibinfo {volume} {50}},\ \bibinfo {pages} {1842} (\bibinfo {year} {2011})}\BibitemShut {NoStop}%
\bibitem [{\citenamefont {Wasko}(2021)}]{3}%
  \BibitemOpen
  \bibfield  {author} {\bibinfo {author} {\bibfnamefont {C.}~\bibnamefont {Wasko}},\ }\bibfield  {title} {\enquote {\bibinfo {title} {Review: Can temperature be used to inform changes to flood extremes with global warming?}}\ }\href {http://dx.doi.org/doi:10.1098/rsta.2019.0551} {\bibfield  {journal} {\bibinfo  {journal} {Phil. Trans. R. Soc. A}\ }\textbf {\bibinfo {volume} {379}},\ \bibinfo {pages} {20190551} (\bibinfo {year} {2021})}\BibitemShut {NoStop}%
\bibitem [{\citenamefont {Ghil}\ \emph {et~al.}(2002)\citenamefont {Ghil}, \citenamefont {Allen}, \citenamefont {Dettinger}, \citenamefont {Ide}, \citenamefont {Kondrashov}, \citenamefont {Mann}, \citenamefont {Robertson}, \citenamefont {Saunders}, \citenamefont {Tian}, \citenamefont {Varadi},\ and\ \citenamefont {Yiou}}]{4}%
  \BibitemOpen
  \bibfield  {author} {\bibinfo {author} {\bibfnamefont {M.}~\bibnamefont {Ghil}}, \bibinfo {author} {\bibfnamefont {M.~R.}\ \bibnamefont {Allen}}, \bibinfo {author} {\bibfnamefont {M.~D.}\ \bibnamefont {Dettinger}}, \bibinfo {author} {\bibfnamefont {K.}~\bibnamefont {Ide}}, \bibinfo {author} {\bibfnamefont {D.}~\bibnamefont {Kondrashov}}, \bibinfo {author} {\bibfnamefont {M.~E.}\ \bibnamefont {Mann}}, \bibinfo {author} {\bibfnamefont {A.~W.}\ \bibnamefont {Robertson}}, \bibinfo {author} {\bibfnamefont {A.}~\bibnamefont {Saunders}}, \bibinfo {author} {\bibfnamefont {Y.}~\bibnamefont {Tian}}, \bibinfo {author} {\bibfnamefont {F.}~\bibnamefont {Varadi}}, \ and\ \bibinfo {author} {\bibfnamefont {P.}~\bibnamefont {Yiou}},\ }\bibfield  {title} {\enquote {\bibinfo {title} {Advanced spectral methods for climatic time series},}\ }\href {http://dx.doi.org/doi:10.1029/2000RG000092} {\bibfield  {journal} {\bibinfo  {journal} {Rev. Geophys.}\ }\textbf {\bibinfo {volume} {40}} (\bibinfo {year} {2002})}\BibitemShut
  {NoStop}%
\bibitem [{\citenamefont {Goswami}\ \emph {et~al.}(2013)\citenamefont {Goswami}, \citenamefont {Marwan}, \citenamefont {Feulner},\ and\ \citenamefont {Kurths}}]{5}%
  \BibitemOpen
  \bibfield  {author} {\bibinfo {author} {\bibfnamefont {B.}~\bibnamefont {Goswami}}, \bibinfo {author} {\bibfnamefont {N.}~\bibnamefont {Marwan}}, \bibinfo {author} {\bibfnamefont {G.}~\bibnamefont {Feulner}}, \ and\ \bibinfo {author} {\bibfnamefont {J.}~\bibnamefont {Kurths}},\ }\bibfield  {title} {\enquote {\bibinfo {title} {How do global temperature drivers influence each other?}}\ }\href {http://dx.doi.org/doi:10.1140/epjst/e2013-01889-8} {\bibfield  {journal} {\bibinfo  {journal} {Eur. Phys. J. Spec. Top.}\ }\textbf {\bibinfo {volume} {222}} (\bibinfo {year} {2013})}\BibitemShut {NoStop}%
\bibitem [{\citenamefont {Ashkenazy}\ \emph {et~al.}(2003)\citenamefont {Ashkenazy}, \citenamefont {Baker}, \citenamefont {Gildor},\ and\ \citenamefont {Havlin}}]{6}%
  \BibitemOpen
  \bibfield  {author} {\bibinfo {author} {\bibfnamefont {Y.}~\bibnamefont {Ashkenazy}}, \bibinfo {author} {\bibfnamefont {D.~R.}\ \bibnamefont {Baker}}, \bibinfo {author} {\bibfnamefont {H.}~\bibnamefont {Gildor}}, \ and\ \bibinfo {author} {\bibfnamefont {S.}~\bibnamefont {Havlin}},\ }\bibfield  {title} {\enquote {\bibinfo {title} {Nonlinearity and multifractality of climate change in the past 420,000 years},}\ }\href {http://dx.doi.org/doi:10.1029/2003GL018099} {\bibfield  {journal} {\bibinfo  {journal} {Geophys. Res. Lett.}\ }\textbf {\bibinfo {volume} {30}},\ \bibinfo {pages} {2146} (\bibinfo {year} {2003})}\BibitemShut {NoStop}%
\bibitem [{\citenamefont {Fountalis}, \citenamefont {Bracco},\ and\ \citenamefont {Dovrolis}(2014)}]{7}%
  \BibitemOpen
  \bibfield  {author} {\bibinfo {author} {\bibfnamefont {I.}~\bibnamefont {Fountalis}}, \bibinfo {author} {\bibfnamefont {A.}~\bibnamefont {Bracco}}, \ and\ \bibinfo {author} {\bibfnamefont {C.}~\bibnamefont {Dovrolis}},\ }\bibfield  {title} {\enquote {\bibinfo {title} {Spatio-temporal network analysis for studying climate patterns},}\ }\href {http://dx.doi.org/doi:10.1007/s00382-013-1729-5} {\bibfield  {journal} {\bibinfo  {journal} {Clim. Dyn.}\ }\textbf {\bibinfo {volume} {42}},\ \bibinfo {pages} {879--899} (\bibinfo {year} {2014})}\BibitemShut {NoStop}%
\bibitem [{\citenamefont {Bradley}\ and\ \citenamefont {Kantz}(2015)}]{8}%
  \BibitemOpen
  \bibfield  {author} {\bibinfo {author} {\bibfnamefont {E.}~\bibnamefont {Bradley}}\ and\ \bibinfo {author} {\bibfnamefont {H.}~\bibnamefont {Kantz}},\ }\bibfield  {title} {\enquote {\bibinfo {title} {Nonlinear time-series analysis revisited},}\ }\href {http://dx.doi.org/doi:10.1063/1.4917289} {\bibfield  {journal} {\bibinfo  {journal} {Chaos}\ }\textbf {\bibinfo {volume} {25}},\ \bibinfo {pages} {097610} (\bibinfo {year} {2015})}\BibitemShut {NoStop}%
\bibitem [{\citenamefont {Ambika}\ and\ \citenamefont {Harikrishnan}()}]{9}%
  \BibitemOpen
  \bibfield  {author} {\bibinfo {author} {\bibfnamefont {G.}~\bibnamefont {Ambika}}\ and\ \bibinfo {author} {\bibfnamefont {K.~P.}\ \bibnamefont {Harikrishnan}},\ }\bibfield  {title} {\enquote {\bibinfo {title} {Methods of nonlinear time series analysis and applications: A review},}\ }\bibfield  {booktitle} {\emph {\bibinfo {booktitle} {Dynamics and control of energy systems}},\ }\href {\doibase 10.1007/978-981-15-0536-2_2} {\ ,\ \bibinfo {pages} {(Springer, 2020), pp. 9--27}}\BibitemShut {NoStop}%
\bibitem [{\citenamefont {Marwan}\ \emph {et~al.}(2007)\citenamefont {Marwan}, \citenamefont {Romano}, \citenamefont {Thiel},\ and\ \citenamefont {Kurths}}]{10}%
  \BibitemOpen
  \bibfield  {author} {\bibinfo {author} {\bibfnamefont {N.}~\bibnamefont {Marwan}}, \bibinfo {author} {\bibfnamefont {M.}~\bibnamefont {Romano}}, \bibinfo {author} {\bibfnamefont {M.}~\bibnamefont {Thiel}}, \ and\ \bibinfo {author} {\bibfnamefont {J.}~\bibnamefont {Kurths}},\ }\bibfield  {title} {\enquote {\bibinfo {title} {Recurrence plots for the analysis of complex systems},}\ }\href {http://dx.doi.org/doi:10.1016/j.physrep.2006.11.001} {\bibfield  {journal} {\bibinfo  {journal} {Phys. Rep.}\ }\textbf {\bibinfo {volume} {438}},\ \bibinfo {pages} {237--329} (\bibinfo {year} {2007})}\BibitemShut {NoStop}%
\bibitem [{\citenamefont {Jacob}\ \emph {et~al.}(2016)\citenamefont {Jacob}, \citenamefont {Harikrishnan}, \citenamefont {Misra},\ and\ \citenamefont {Ambika}}]{11}%
  \BibitemOpen
  \bibfield  {author} {\bibinfo {author} {\bibfnamefont {R.}~\bibnamefont {Jacob}}, \bibinfo {author} {\bibfnamefont {K.~P.}\ \bibnamefont {Harikrishnan}}, \bibinfo {author} {\bibfnamefont {R.}~\bibnamefont {Misra}}, \ and\ \bibinfo {author} {\bibfnamefont {G.}~\bibnamefont {Ambika}},\ }\bibfield  {title} {\enquote {\bibinfo {title} {Uniform framework for the recurrence-network analysis of chaotic time series},}\ }\href {http://dx.doi.org/doi:10.1103/PhysRevE.93.012202} {\bibfield  {journal} {\bibinfo  {journal} {Phys. Rev. E}\ }\textbf {\bibinfo {volume} {93}},\ \bibinfo {pages} {012202} (\bibinfo {year} {2016})}\BibitemShut {NoStop}%
\bibitem [{\citenamefont {Marwan}\ \emph {et~al.}(2013)\citenamefont {Marwan} \emph {et~al.}}]{12}%
  \BibitemOpen
  \bibfield  {author} {\bibinfo {author} {\bibfnamefont {N.}~\bibnamefont {Marwan}} \emph {et~al.},\ }\bibfield  {title} {\enquote {\bibinfo {title} {Recurrence plots 25 years later -gaining confidence in dynamical transitions},}\ }\href {http://dx.doi.org/doi:10.1209/0295-5075/101/20007} {\bibfield  {journal} {\bibinfo  {journal} {Europhys. Lett.}\ }\textbf {\bibinfo {volume} {101}},\ \bibinfo {pages} {20007} (\bibinfo {year} {2013})}\BibitemShut {NoStop}%
\bibitem [{\citenamefont {Goswami}(2019)}]{13}%
  \BibitemOpen
  \bibfield  {author} {\bibinfo {author} {\bibfnamefont {B.}~\bibnamefont {Goswami}},\ }\bibfield  {title} {\enquote {\bibinfo {title} {A brief introduction to nonlinear time series analysis and recurrence plots},}\ }\href {http://dx.doi.org/doi:10.3390/vibration2040021} {\bibfield  {journal} {\bibinfo  {journal} {Vibration}\ }\textbf {\bibinfo {volume} {2}},\ \bibinfo {pages} {332--368} (\bibinfo {year} {2019})}\BibitemShut {NoStop}%
\bibitem [{\citenamefont {Lekscha}\ and\ \citenamefont {Donner}(2020)}]{14}%
  \BibitemOpen
  \bibfield  {author} {\bibinfo {author} {\bibfnamefont {J.}~\bibnamefont {Lekscha}}\ and\ \bibinfo {author} {\bibfnamefont {R.~V.}\ \bibnamefont {Donner}},\ }\bibfield  {title} {\enquote {\bibinfo {title} {Detecting dynamical anomalies in time series from different palaeoclimate proxy archives using windowed recurrence network analysis},}\ }\href {http://dx.doi.org/doi:10.5194/npg-27-261-2020} {\bibfield  {journal} {\bibinfo  {journal} {Nonlinear Processes Geophys.}\ }\textbf {\bibinfo {volume} {27}},\ \bibinfo {pages} {261--275} (\bibinfo {year} {2020})}\BibitemShut {NoStop}%
\bibitem [{\citenamefont {Donges}\ \emph {et~al.}(2011)\citenamefont {Donges}, \citenamefont {Donner}, \citenamefont {Rehfeld}, \citenamefont {Marwan}, \citenamefont {Trauth},\ and\ \citenamefont {Kurths}}]{15}%
  \BibitemOpen
  \bibfield  {author} {\bibinfo {author} {\bibfnamefont {J.~F.}\ \bibnamefont {Donges}}, \bibinfo {author} {\bibfnamefont {R.~V.}\ \bibnamefont {Donner}}, \bibinfo {author} {\bibfnamefont {K.}~\bibnamefont {Rehfeld}}, \bibinfo {author} {\bibfnamefont {N.}~\bibnamefont {Marwan}}, \bibinfo {author} {\bibfnamefont {M.}~\bibnamefont {Trauth}}, \ and\ \bibinfo {author} {\bibfnamefont {J.}~\bibnamefont {Kurths}},\ }\bibfield  {title} {\enquote {\bibinfo {title} {Identification of dynamical transitions in marine palaeoclimate records by recurrence network analysis},}\ }\href {http://dx.doi.org/doi:10.5194/npg-18-545-2011} {\bibfield  {journal} {\bibinfo  {journal} {Nonlinear Processes Geophys.}\ }\textbf {\bibinfo {volume} {18}},\ \bibinfo {pages} {545--562} (\bibinfo {year} {2011})}\BibitemShut {NoStop}%
\bibitem [{\citenamefont {George}, \citenamefont {Misra},\ and\ \citenamefont {Ambika}(2019)}]{16}%
  \BibitemOpen
  \bibfield  {author} {\bibinfo {author} {\bibfnamefont {S.~V.}\ \bibnamefont {George}}, \bibinfo {author} {\bibfnamefont {R.}~\bibnamefont {Misra}}, \ and\ \bibinfo {author} {\bibfnamefont {G.}~\bibnamefont {Ambika}},\ }\bibfield  {title} {\enquote {\bibinfo {title} {Classification of close binary stars using recurrence networks},}\ }\href {http://dx.doi.org/doi:10.1063/1.5120739} {\bibfield  {journal} {\bibinfo  {journal} {Chaos}\ }\textbf {\bibinfo {volume} {29}},\ \bibinfo {pages} {113112} (\bibinfo {year} {2019})}\BibitemShut {NoStop}%
\bibitem [{\citenamefont {Godavarthi}\ \emph {et~al.}(2017)\citenamefont {Godavarthi}, \citenamefont {Unni}, \citenamefont {Gopalakrishnan},\ and\ \citenamefont {Sujith}}]{17}%
  \BibitemOpen
  \bibfield  {author} {\bibinfo {author} {\bibfnamefont {V.}~\bibnamefont {Godavarthi}}, \bibinfo {author} {\bibfnamefont {V.}~\bibnamefont {Unni}}, \bibinfo {author} {\bibfnamefont {E.}~\bibnamefont {Gopalakrishnan}}, \ and\ \bibinfo {author} {\bibfnamefont {R.}~\bibnamefont {Sujith}},\ }\bibfield  {title} {\enquote {\bibinfo {title} {Recurrence networks to study dynamical transitions in a turbulent combustor},}\ }\href {http://dx.doi.org/doi:10.1063/1.4985275} {\bibfield  {journal} {\bibinfo  {journal} {Chaos}\ }\textbf {\bibinfo {volume} {27}},\ \bibinfo {pages} {063113} (\bibinfo {year} {2017})}\BibitemShut {NoStop}%
\bibitem [{\citenamefont {Bhattacharya}\ \emph {et~al.}(2022)\citenamefont {Bhattacharya}, \citenamefont {De}, \citenamefont {Mondal}, \citenamefont {Mukhopadhyay},\ and\ \citenamefont {Sen}}]{18}%
  \BibitemOpen
  \bibfield  {author} {\bibinfo {author} {\bibfnamefont {A.}~\bibnamefont {Bhattacharya}}, \bibinfo {author} {\bibfnamefont {S.}~\bibnamefont {De}}, \bibinfo {author} {\bibfnamefont {S.}~\bibnamefont {Mondal}}, \bibinfo {author} {\bibfnamefont {A.}~\bibnamefont {Mukhopadhyay}}, \ and\ \bibinfo {author} {\bibfnamefont {S.}~\bibnamefont {Sen}},\ }\bibfield  {title} {\enquote {\bibinfo {title} {Early detection of lean blowout using recurrence network for varying degrees of premixedness},}\ }\href {http://dx.doi.org/doi:10.1063/5.0077436} {\bibfield  {journal} {\bibinfo  {journal} {Chaos}\ }\textbf {\bibinfo {volume} {32}},\ \bibinfo {pages} {063105} (\bibinfo {year} {2022})}\BibitemShut {NoStop}%
\bibitem [{\citenamefont {Mart{\'\i}n-Gonz{\'a}lez}\ \emph {et~al.}(2018)\citenamefont {Mart{\'\i}n-Gonz{\'a}lez}, \citenamefont {Navarro-Mesa}, \citenamefont {Juli{\'a}-Serd{\'a}}, \citenamefont {Ram{\'\i}rez-{\'A}vila},\ and\ \citenamefont {Ravelo-Garc{\'\i}a}}]{19}%
  \BibitemOpen
  \bibfield  {author} {\bibinfo {author} {\bibfnamefont {S.}~\bibnamefont {Mart{\'\i}n-Gonz{\'a}lez}}, \bibinfo {author} {\bibfnamefont {J.}~\bibnamefont {Navarro-Mesa}}, \bibinfo {author} {\bibfnamefont {G.}~\bibnamefont {Juli{\'a}-Serd{\'a}}}, \bibinfo {author} {\bibfnamefont {G.}~\bibnamefont {Ram{\'\i}rez-{\'A}vila}}, \ and\ \bibinfo {author} {\bibfnamefont {A.}~\bibnamefont {Ravelo-Garc{\'\i}a}},\ }\bibfield  {title} {\enquote {\bibinfo {title} {Improving the understanding of sleep apnea characterization using recurrence quantification analysis by defining overall acceptable values for the dimensionality of the system, the delay, and the distance threshold},}\ }\href {http://dx.doi.org/doi:10.1371/journal.pone.0194462} {\bibfield  {journal} {\bibinfo  {journal} {PLoS One}\ }\textbf {\bibinfo {volume} {13}},\ \bibinfo {pages} {e0194462} (\bibinfo {year} {2018})}\BibitemShut {NoStop}%
\bibitem [{\citenamefont {Krishnadas}, \citenamefont {Harikrishnan},\ and\ \citenamefont {Ambika}(2022)}]{20}%
  \BibitemOpen
  \bibfield  {author} {\bibinfo {author} {\bibfnamefont {M.}~\bibnamefont {Krishnadas}}, \bibinfo {author} {\bibfnamefont {K.~P.}\ \bibnamefont {Harikrishnan}}, \ and\ \bibinfo {author} {\bibfnamefont {G.}~\bibnamefont {Ambika}},\ }\bibfield  {title} {\enquote {\bibinfo {title} {Recurrence measures and transitions in stock market dynamics},}\ }\href {http://dx.doi.org/doi:10.1016/j.physa.2022.128240} {\bibfield  {journal} {\bibinfo  {journal} {Physica A}\ }\textbf {\bibinfo {volume} {608}},\ \bibinfo {pages} {128240} (\bibinfo {year} {2022})}\BibitemShut {NoStop}%
\bibitem [{\citenamefont {George}\ \emph {et~al.}(2020)\citenamefont {George}, \citenamefont {Kachhara}, \citenamefont {Misra},\ and\ \citenamefont {Ambika}}]{21}%
  \BibitemOpen
  \bibfield  {author} {\bibinfo {author} {\bibfnamefont {S.~V.}\ \bibnamefont {George}}, \bibinfo {author} {\bibfnamefont {S.}~\bibnamefont {Kachhara}}, \bibinfo {author} {\bibfnamefont {R.}~\bibnamefont {Misra}}, \ and\ \bibinfo {author} {\bibfnamefont {G.}~\bibnamefont {Ambika}},\ }\bibfield  {title} {\enquote {\bibinfo {title} {Early warning signals indicate a critical transition in betelgeuse},}\ }\href {http://dx.doi.org/doi:10.1051/0004-6361/202038785} {\bibfield  {journal} {\bibinfo  {journal} {Astron. Astrophys.}\ }\textbf {\bibinfo {volume} {640}},\ \bibinfo {pages} {L21} (\bibinfo {year} {2020})}\BibitemShut {NoStop}%
\bibitem [{\citenamefont {Kachhara}\ and\ \citenamefont {Ambika}(2019)}]{22}%
  \BibitemOpen
  \bibfield  {author} {\bibinfo {author} {\bibfnamefont {S.}~\bibnamefont {Kachhara}}\ and\ \bibinfo {author} {\bibfnamefont {G.}~\bibnamefont {Ambika}},\ }\bibfield  {title} {\enquote {\bibinfo {title} {Bimodality and scaling in recurrence networks from ecg data},}\ }\href {http://dx.doi.org/doi:10.1209/0295-5075/127/60004} {\bibfield  {journal} {\bibinfo  {journal} {Europhys. Lett.}\ }\textbf {\bibinfo {volume} {127}},\ \bibinfo {pages} {60004} (\bibinfo {year} {2019})}\BibitemShut {NoStop}%
\bibitem [{\citenamefont {Adarsh}\ \emph {et~al.}(2020)\citenamefont {Adarsh}, \citenamefont {Nourani}, \citenamefont {Archana},\ and\ \citenamefont {Dharan}}]{23}%
  \BibitemOpen
  \bibfield  {author} {\bibinfo {author} {\bibfnamefont {S.}~\bibnamefont {Adarsh}}, \bibinfo {author} {\bibfnamefont {V.}~\bibnamefont {Nourani}}, \bibinfo {author} {\bibfnamefont {D.}~\bibnamefont {Archana}}, \ and\ \bibinfo {author} {\bibfnamefont {D.~S.}\ \bibnamefont {Dharan}},\ }\bibfield  {title} {\enquote {\bibinfo {title} {Multifractal description of daily rainfall fields over india},}\ }\href {http://dx.doi.org/doi:10.1016/j.jhydrol.2020.124913} {\bibfield  {journal} {\bibinfo  {journal} {J. Hydrol.}\ }\textbf {\bibinfo {volume} {586}},\ \bibinfo {pages} {124913} (\bibinfo {year} {2020})}\BibitemShut {NoStop}%
\bibitem [{\citenamefont {Ray}\ \emph {et~al.}(2018)\citenamefont {Ray}, \citenamefont {Dey}, \citenamefont {Khondekar},\ and\ \citenamefont {Ghosh}}]{24}%
  \BibitemOpen
  \bibfield  {author} {\bibinfo {author} {\bibfnamefont {R.}~\bibnamefont {Ray}}, \bibinfo {author} {\bibfnamefont {S.}~\bibnamefont {Dey}}, \bibinfo {author} {\bibfnamefont {M.~H.}\ \bibnamefont {Khondekar}}, \ and\ \bibinfo {author} {\bibfnamefont {K.}~\bibnamefont {Ghosh}},\ }\bibfield  {title} {\enquote {\bibinfo {title} {Multifractality and singularity in average temperature and dew point across india},}\ }\href {http://dx.doi.org/doi:10.19101/IJATEE.2018.542018} {\bibfield  {journal} {\bibinfo  {journal} {Int. J. Adv. Technol. Eng. Explor.}\ }\textbf {\bibinfo {volume} {5}},\ \bibinfo {pages} {2394--5443} (\bibinfo {year} {2018})}\BibitemShut {NoStop}%
\bibitem [{\citenamefont {Hingane}, \citenamefont {Kumar},\ and\ \citenamefont {Murty}(1985)}]{25}%
  \BibitemOpen
  \bibfield  {author} {\bibinfo {author} {\bibfnamefont {L.}~\bibnamefont {Hingane}}, \bibinfo {author} {\bibfnamefont {K.}~\bibnamefont {Kumar}}, \ and\ \bibinfo {author} {\bibfnamefont {B.}~\bibnamefont {Murty}},\ }\bibfield  {title} {\enquote {\bibinfo {title} {Long term trends of surface air temperature in india},}\ }\href {http://dx.doi.org/doi:10.1002/joc.3370050505} {\bibfield  {journal} {\bibinfo  {journal} {J. Clim.}\ }\textbf {\bibinfo {volume} {5}},\ \bibinfo {pages} {521--528} (\bibinfo {year} {1985})}\BibitemShut {NoStop}%
\bibitem [{\citenamefont {Paul}, \citenamefont {Birthal},\ and\ \citenamefont {Khokhar}(2014)}]{26}%
  \BibitemOpen
  \bibfield  {author} {\bibinfo {author} {\bibfnamefont {R.~K.}\ \bibnamefont {Paul}}, \bibinfo {author} {\bibfnamefont {P.}~\bibnamefont {Birthal}}, \ and\ \bibinfo {author} {\bibfnamefont {A.}~\bibnamefont {Khokhar}},\ }\bibfield  {title} {\enquote {\bibinfo {title} {Structural breaks in mean temperature over agroclimatic zones in india},}\ }\href {\doibase 10.1155/2014/434325} {\bibfield  {journal} {\bibinfo  {journal} {Sci. World J.}\ }\textbf {\bibinfo {volume} {2014}},\ \bibinfo {pages} {434325} (\bibinfo {year} {2014})}\BibitemShut {NoStop}%
\bibitem [{\citenamefont {Rao}, \citenamefont {Murty},\ and\ \citenamefont {Joshi}(2005)}]{27}%
  \BibitemOpen
  \bibfield  {author} {\bibinfo {author} {\bibfnamefont {G.}~\bibnamefont {Rao}}, \bibinfo {author} {\bibfnamefont {M.}~\bibnamefont {Murty}}, \ and\ \bibinfo {author} {\bibfnamefont {U.}~\bibnamefont {Joshi}},\ }\bibfield  {title} {\enquote {\bibinfo {title} {Climate change over india as revealed by critical extreme temperature analysis},}\ }\href {http://dx.doi.org/doi:10.54302/mausam.v56i3.990} {\bibfield  {journal} {\bibinfo  {journal} {Mausam}\ }\textbf {\bibinfo {volume} {56}},\ \bibinfo {pages} {601--608} (\bibinfo {year} {2005})}\BibitemShut {NoStop}%
\bibitem [{\citenamefont {Beck}\ \emph {et~al.}(2018)\citenamefont {Beck}, \citenamefont {Zimmermann}, \citenamefont {McVicar} \emph {et~al.}}]{28}%
  \BibitemOpen
  \bibfield  {author} {\bibinfo {author} {\bibfnamefont {H.}~\bibnamefont {Beck}}, \bibinfo {author} {\bibfnamefont {N.}~\bibnamefont {Zimmermann}}, \bibinfo {author} {\bibfnamefont {T.}~\bibnamefont {McVicar}},  \emph {et~al.},\ }\bibfield  {title} {\enquote {\bibinfo {title} {Present and future köppen-geiger climate classification maps at 1-km resolution},}\ }\href {\doibase 10.1038/sdata.2018.214} {\bibfield  {journal} {\bibinfo  {journal} {Sci. Data}\ }\textbf {\bibinfo {volume} {5}},\ \bibinfo {pages} {180214} (\bibinfo {year} {2018})}\BibitemShut {NoStop}%
\bibitem [{\citenamefont {Kraemer}\ \emph {et~al.}(2018)\citenamefont {Kraemer}, \citenamefont {Donner}, \citenamefont {Heitzig},\ and\ \citenamefont {Marwan}}]{rcthreshold}%
  \BibitemOpen
  \bibfield  {author} {\bibinfo {author} {\bibfnamefont {K.~H.}\ \bibnamefont {Kraemer}}, \bibinfo {author} {\bibfnamefont {R.}~\bibnamefont {Donner}}, \bibinfo {author} {\bibfnamefont {J.}~\bibnamefont {Heitzig}}, \ and\ \bibinfo {author} {\bibfnamefont {N.}~\bibnamefont {Marwan}},\ }\bibfield  {title} {\enquote {\bibinfo {title} {Recurrence threshold selection for obtaining robust recurrence characteristics in different embedding dimensions},}\ }\href {\doibase 10.1063/1.5024914} {\bibfield  {journal} {\bibinfo  {journal} {Chaos}\ }\textbf {\bibinfo {volume} {28}} (\bibinfo {year} {2018}),\ 10.1063/1.5024914}\BibitemShut {NoStop}%
\bibitem [{\citenamefont {Song}, \citenamefont {Yeh},\ and\ \citenamefont {Park}(2020)}]{29}%
  \BibitemOpen
  \bibfield  {author} {\bibinfo {author} {\bibfnamefont {S.}~\bibnamefont {Song}}, \bibinfo {author} {\bibfnamefont {S.}~\bibnamefont {Yeh}}, \ and\ \bibinfo {author} {\bibfnamefont {J.}~\bibnamefont {Park}},\ }\bibfield  {title} {\enquote {\bibinfo {title} {Dissimilar characteristics associated with the 1976/1977 and 1998/1999 climate regime shifts in the north pacific},}\ }\href {http://dx.doi.org/doi:10.1007/s00704-020-03378-y} {\bibfield  {journal} {\bibinfo  {journal} {Theor. Appl. Clim.}\ }\textbf {\bibinfo {volume} {142}},\ \bibinfo {pages} {1463--1470} (\bibinfo {year} {2020})}\BibitemShut {NoStop}%
\bibitem [{\citenamefont {Hong}\ \emph {et~al.}(2014{\natexlab{a}})\citenamefont {Hong}, \citenamefont {Wu}, \citenamefont {Li},\ and\ \citenamefont {Chang}}]{30}%
  \BibitemOpen
  \bibfield  {author} {\bibinfo {author} {\bibfnamefont {C.-C.}\ \bibnamefont {Hong}}, \bibinfo {author} {\bibfnamefont {Y.-K.}\ \bibnamefont {Wu}}, \bibinfo {author} {\bibfnamefont {T.}~\bibnamefont {Li}}, \ and\ \bibinfo {author} {\bibfnamefont {C.-C.}\ \bibnamefont {Chang}},\ }\bibfield  {title} {\enquote {\bibinfo {title} {The climate regime shift over the pacific during 1996/1997},}\ }\href {http://dx.doi.org/doi:10.1007/s00382-013-1867-9} {\bibfield  {journal} {\bibinfo  {journal} {Clim. Dyn.}\ }\textbf {\bibinfo {volume} {43}},\ \bibinfo {pages} {435--446} (\bibinfo {year} {2014}{\natexlab{a}})}\BibitemShut {NoStop}%
\bibitem [{\citenamefont {Nitta}\ and\ \citenamefont {Yamada}(1989)}]{31}%
  \BibitemOpen
  \bibfield  {author} {\bibinfo {author} {\bibfnamefont {T.}~\bibnamefont {Nitta}}\ and\ \bibinfo {author} {\bibfnamefont {S.}~\bibnamefont {Yamada}},\ }\bibfield  {title} {\enquote {\bibinfo {title} {Recent warming of tropical sea surface temperature and its relationship to the northern hemisphere circulation},}\ }\href {http://dx.doi.org/doi:10.2151/jmsj1965.67.3_375} {\bibfield  {journal} {\bibinfo  {journal} {J. Meteorol. Soc. Jpn.}\ }\textbf {\bibinfo {volume} {67}},\ \bibinfo {pages} {375--383} (\bibinfo {year} {1989})}\BibitemShut {NoStop}%
\bibitem [{\citenamefont {Hong}\ \emph {et~al.}(2014{\natexlab{b}})\citenamefont {Hong}, \citenamefont {Wu}, \citenamefont {Li},\ and\ \citenamefont {Chang}}]{32}%
  \BibitemOpen
  \bibfield  {author} {\bibinfo {author} {\bibfnamefont {C.-C.}\ \bibnamefont {Hong}}, \bibinfo {author} {\bibfnamefont {Y.-K.}\ \bibnamefont {Wu}}, \bibinfo {author} {\bibfnamefont {T.}~\bibnamefont {Li}}, \ and\ \bibinfo {author} {\bibfnamefont {C.-C.}\ \bibnamefont {Chang}},\ }\bibfield  {title} {\enquote {\bibinfo {title} {The climate regime shift over the pacific during 1996/1997},}\ }\href {\doibase 10.1007/s00382-013-1867-9} {\bibfield  {journal} {\bibinfo  {journal} {Clim. Dyn.}\ }\textbf {\bibinfo {volume} {43}},\ \bibinfo {pages} {435--446} (\bibinfo {year} {2014}{\natexlab{b}})}\BibitemShut {NoStop}%
\bibitem [{\citenamefont {Jo}, \citenamefont {Yeh},\ and\ \citenamefont {Lee}(2015)}]{33}%
  \BibitemOpen
  \bibfield  {author} {\bibinfo {author} {\bibfnamefont {H.-S.}\ \bibnamefont {Jo}}, \bibinfo {author} {\bibfnamefont {S.-W.}\ \bibnamefont {Yeh}}, \ and\ \bibinfo {author} {\bibfnamefont {S.-K.}\ \bibnamefont {Lee}},\ }\bibfield  {title} {\enquote {\bibinfo {title} {Changes in the relationship in the sst variability between the tropical pacific and the north pacific across the 1998/1999 regime shift},}\ }\href {http://dx.doi.org/doi:10.1002/2015GL065049} {\bibfield  {journal} {\bibinfo  {journal} {Geophys. Res. Lett.}\ }\textbf {\bibinfo {volume} {42}},\ \bibinfo {pages} {7171--7178} (\bibinfo {year} {2015})}\BibitemShut {NoStop}%
\bibitem [{\citenamefont {Miller}\ \emph {et~al.}(1994)\citenamefont {Miller}, \citenamefont {Cayan}, \citenamefont {Barnett}, \citenamefont {Graham},\ and\ \citenamefont {Oberhuber}}]{34}%
  \BibitemOpen
  \bibfield  {author} {\bibinfo {author} {\bibfnamefont {A.}~\bibnamefont {Miller}}, \bibinfo {author} {\bibfnamefont {D.}~\bibnamefont {Cayan}}, \bibinfo {author} {\bibfnamefont {T.}~\bibnamefont {Barnett}}, \bibinfo {author} {\bibfnamefont {N.}~\bibnamefont {Graham}}, \ and\ \bibinfo {author} {\bibfnamefont {J.}~\bibnamefont {Oberhuber}},\ }\bibfield  {title} {\enquote {\bibinfo {title} {The 1976-77 climate shift of the pacific ocean},}\ }\href {http://dx.doi.org/doi:10.5670/oceanog.1994.11} {\bibfield  {journal} {\bibinfo  {journal} {Oceanography}\ }\textbf {\bibinfo {volume} {7}},\ \bibinfo {pages} {21--26} (\bibinfo {year} {1994})}\BibitemShut {NoStop}%
\bibitem [{\citenamefont {M{\"u}llner}(2011)}]{spearman1}%
  \BibitemOpen
  \bibfield  {author} {\bibinfo {author} {\bibfnamefont {D.}~\bibnamefont {M{\"u}llner}},\ }\bibfield  {title} {\enquote {\bibinfo {title} {Modern hierarchical, agglomerative clustering algorithms},}\ }\href {https://arxiv.org/abs/1109.2378} {\bibfield  {journal} {\bibinfo  {journal} {arXiv preprint arXiv:1109.2378}\ } (\bibinfo {year} {2011})}\BibitemShut {NoStop}%
\bibitem [{\citenamefont {Bar-Joseph}, \citenamefont {Gifford},\ and\ \citenamefont {Jaakkola}(2001)}]{spearman2}%
  \BibitemOpen
  \bibfield  {author} {\bibinfo {author} {\bibfnamefont {Z.}~\bibnamefont {Bar-Joseph}}, \bibinfo {author} {\bibfnamefont {D.~K.}\ \bibnamefont {Gifford}}, \ and\ \bibinfo {author} {\bibfnamefont {T.~S.}\ \bibnamefont {Jaakkola}},\ }\bibfield  {title} {\enquote {\bibinfo {title} {Fast optimal leaf ordering for hierarchical clustering},}\ }\href {http://dx.doi.org/doi:10.1093/bioinformatics/17.suppl_1.S22} {\bibfield  {journal} {\bibinfo  {journal} {Bioinformatics}\ }\textbf {\bibinfo {volume} {17}},\ \bibinfo {pages} {S22--S29} (\bibinfo {year} {2001})}\BibitemShut {NoStop}%
\bibitem [{\citenamefont {Rousseeuw}(1987)}]{silhouette}%
  \BibitemOpen
  \bibfield  {author} {\bibinfo {author} {\bibfnamefont {P.~J.}\ \bibnamefont {Rousseeuw}},\ }\bibfield  {title} {\enquote {\bibinfo {title} {Silhouettes: a graphical aid to the interpretation and validation of cluster analysis},}\ }\href {http://dx.doi.org/doi:10.1016/0377-0427(87)90125-7} {\bibfield  {journal} {\bibinfo  {journal} {J. Comput. Appl. Math.}\ }\textbf {\bibinfo {volume} {20}},\ \bibinfo {pages} {53--65} (\bibinfo {year} {1987})}\BibitemShut {NoStop}%
\bibitem [{\citenamefont {Braun}\ \emph {et~al.}(2021)\citenamefont {Braun}, \citenamefont {Unni}, \citenamefont {Sujith}, \citenamefont {Kurths},\ and\ \citenamefont {Marwan}}]{tobias}%
  \BibitemOpen
  \bibfield  {author} {\bibinfo {author} {\bibfnamefont {T.}~\bibnamefont {Braun}}, \bibinfo {author} {\bibfnamefont {V.~R.}\ \bibnamefont {Unni}}, \bibinfo {author} {\bibfnamefont {R.}~\bibnamefont {Sujith}}, \bibinfo {author} {\bibfnamefont {J.}~\bibnamefont {Kurths}}, \ and\ \bibinfo {author} {\bibfnamefont {N.}~\bibnamefont {Marwan}},\ }\bibfield  {title} {\enquote {\bibinfo {title} {Detection of dynamical regime transitions with lacunarity as a multiscale recurrence quantification measure},}\ }\href {http://dx.doi.org/doi:10.1007/s11071-021-06457-5} {\bibfield  {journal} {\bibinfo  {journal} {Nonlinear Dyn.}\ }\textbf {\bibinfo {volume} {104}},\ \bibinfo {pages} {3955--3973} (\bibinfo {year} {2021})}\BibitemShut {NoStop}%
\bibitem [{\citenamefont {Roxy}\ \emph {et~al.}(2015)\citenamefont {Roxy}, \citenamefont {Ritika}, \citenamefont {Terray}, \citenamefont {Murtugudde}, \citenamefont {Ashok},\ and\ \citenamefont {Goswami}}]{40}%
  \BibitemOpen
  \bibfield  {author} {\bibinfo {author} {\bibfnamefont {M.~K.}\ \bibnamefont {Roxy}}, \bibinfo {author} {\bibfnamefont {K.}~\bibnamefont {Ritika}}, \bibinfo {author} {\bibfnamefont {P.}~\bibnamefont {Terray}}, \bibinfo {author} {\bibfnamefont {R.}~\bibnamefont {Murtugudde}}, \bibinfo {author} {\bibfnamefont {K.}~\bibnamefont {Ashok}}, \ and\ \bibinfo {author} {\bibfnamefont {B.}~\bibnamefont {Goswami}},\ }\bibfield  {title} {\enquote {\bibinfo {title} {Drying of indian subcontinent by rapid indian ocean warming and a weakening land-sea thermal gradient},}\ }\href {http://dx.doi.org/doi:10.1038/ncomms8423} {\bibfield  {journal} {\bibinfo  {journal} {Nat. Commun.}\ }\textbf {\bibinfo {volume} {6}},\ \bibinfo {pages} {7423} (\bibinfo {year} {2015})}\BibitemShut {NoStop}%
\bibitem [{\citenamefont {Sahana}\ \emph {et~al.}(2015{\natexlab{a}})\citenamefont {Sahana}, \citenamefont {Ghosh}, \citenamefont {Ganguly},\ and\ \citenamefont {Murtugudde}}]{41}%
  \BibitemOpen
  \bibfield  {author} {\bibinfo {author} {\bibfnamefont {A.}~\bibnamefont {Sahana}}, \bibinfo {author} {\bibfnamefont {S.}~\bibnamefont {Ghosh}}, \bibinfo {author} {\bibfnamefont {A.}~\bibnamefont {Ganguly}}, \ and\ \bibinfo {author} {\bibfnamefont {R.}~\bibnamefont {Murtugudde}},\ }\bibfield  {title} {\enquote {\bibinfo {title} {Shift in indian summer monsoon onset during 1976/1977},}\ }\href {http://dx.doi.org/doi:10.1088/1748-9326/10/5/054006} {\bibfield  {journal} {\bibinfo  {journal} {Environ. Res. Lett.}\ }\textbf {\bibinfo {volume} {10}},\ \bibinfo {pages} {054006} (\bibinfo {year} {2015}{\natexlab{a}})}\BibitemShut {NoStop}%
\bibitem [{\citenamefont {Menon}\ \emph {et~al.}(2002)\citenamefont {Menon}, \citenamefont {Hansen}, \citenamefont {Nazarenko},\ and\ \citenamefont {Luo}}]{42}%
  \BibitemOpen
  \bibfield  {author} {\bibinfo {author} {\bibfnamefont {S.}~\bibnamefont {Menon}}, \bibinfo {author} {\bibfnamefont {J.}~\bibnamefont {Hansen}}, \bibinfo {author} {\bibfnamefont {L.}~\bibnamefont {Nazarenko}}, \ and\ \bibinfo {author} {\bibfnamefont {Y.}~\bibnamefont {Luo}},\ }\bibfield  {title} {\enquote {\bibinfo {title} {Climate effects of black carbon aerosols in china and india},}\ }\href {http://dx.doi.org/doi:10.1126/science.1075159} {\bibfield  {journal} {\bibinfo  {journal} {Science}\ }\textbf {\bibinfo {volume} {297}},\ \bibinfo {pages} {2250--2253} (\bibinfo {year} {2002})}\BibitemShut {NoStop}%
\bibitem [{\citenamefont {Prasad}, \citenamefont {Singh},\ and\ \citenamefont {Singh}(2004)}]{43}%
  \BibitemOpen
  \bibfield  {author} {\bibinfo {author} {\bibfnamefont {A.~K.}\ \bibnamefont {Prasad}}, \bibinfo {author} {\bibfnamefont {R.~P.}\ \bibnamefont {Singh}}, \ and\ \bibinfo {author} {\bibfnamefont {A.}~\bibnamefont {Singh}},\ }\bibfield  {title} {\enquote {\bibinfo {title} {Variability of aerosol optical depth over indian subcontinent using modis data},}\ }\href {http://dx.doi.org/doi:10.1007/BF03030855} {\bibfield  {journal} {\bibinfo  {journal} {J. Indian Soc. Remote Sens.}\ }\textbf {\bibinfo {volume} {32}},\ \bibinfo {pages} {313--316} (\bibinfo {year} {2004})}\BibitemShut {NoStop}%
\bibitem [{\citenamefont {Zhu}\ and\ \citenamefont {Houghton}(1996)}]{44}%
  \BibitemOpen
  \bibfield  {author} {\bibinfo {author} {\bibfnamefont {Y.}~\bibnamefont {Zhu}}\ and\ \bibinfo {author} {\bibfnamefont {D.~D.}\ \bibnamefont {Houghton}},\ }\bibfield  {title} {\enquote {\bibinfo {title} {The impact of indian ocean sst on the large-scale asian summer monsoon and the hydrological cycle},}\ }\href {http://dx.doi.org/doi:10.1002/(SICI)1097-0088(199606)16:6<617::AID-JOC32>3.0.CO;2-I} {\bibfield  {journal} {\bibinfo  {journal} {Int. J. Climatol.: A J. R. Meteorol. Soc.}\ }\textbf {\bibinfo {volume} {16}},\ \bibinfo {pages} {617--632} (\bibinfo {year} {1996})}\BibitemShut {NoStop}%
\bibitem [{\citenamefont {Roy}, \citenamefont {Tedeschi},\ and\ \citenamefont {Collins}(2019)}]{45}%
  \BibitemOpen
  \bibfield  {author} {\bibinfo {author} {\bibfnamefont {I.}~\bibnamefont {Roy}}, \bibinfo {author} {\bibfnamefont {R.~G.}\ \bibnamefont {Tedeschi}}, \ and\ \bibinfo {author} {\bibfnamefont {M.}~\bibnamefont {Collins}},\ }\bibfield  {title} {\enquote {\bibinfo {title} {Enso teleconnections to the indian summer monsoon under changing climate},}\ }\href {http://dx.doi.org/doi:10.1002/joc.5999} {\bibfield  {journal} {\bibinfo  {journal} {Int. J. Climatol.}\ }\textbf {\bibinfo {volume} {39}},\ \bibinfo {pages} {3031--3042} (\bibinfo {year} {2019})}\BibitemShut {NoStop}%
\bibitem [{\citenamefont {Saikranthi}\ \emph {et~al.}(2018)\citenamefont {Saikranthi}, \citenamefont {Radhakrishna}, \citenamefont {Satheesh} \emph {et~al.}}]{39}%
  \BibitemOpen
  \bibfield  {author} {\bibinfo {author} {\bibfnamefont {K.}~\bibnamefont {Saikranthi}}, \bibinfo {author} {\bibfnamefont {B.}~\bibnamefont {Radhakrishna}}, \bibinfo {author} {\bibfnamefont {S.~K.}\ \bibnamefont {Satheesh}},  \emph {et~al.},\ }\bibfield  {title} {\enquote {\bibinfo {title} {Spatial variation of different rain systems during el niño and la niña periods over india and adjoining ocean},}\ }\href {\doibase 10.1007/s00382-017-3833-4} {\bibfield  {journal} {\bibinfo  {journal} {Clim. Dyn.}\ }\textbf {\bibinfo {volume} {50}},\ \bibinfo {pages} {3671–3685} (\bibinfo {year} {2018})}\BibitemShut {NoStop}%
\bibitem [{\citenamefont {Athira}\ \emph {et~al.}(2023)\citenamefont {Athira}, \citenamefont {Roxy}, \citenamefont {Dasgupta}, \citenamefont {Saranya}, \citenamefont {Singh},\ and\ \citenamefont {Attada}}]{46}%
  \BibitemOpen
  \bibfield  {author} {\bibinfo {author} {\bibfnamefont {K.}~\bibnamefont {Athira}}, \bibinfo {author} {\bibfnamefont {M.~K.}\ \bibnamefont {Roxy}}, \bibinfo {author} {\bibfnamefont {P.}~\bibnamefont {Dasgupta}}, \bibinfo {author} {\bibfnamefont {J.}~\bibnamefont {Saranya}}, \bibinfo {author} {\bibfnamefont {V.~K.}\ \bibnamefont {Singh}}, \ and\ \bibinfo {author} {\bibfnamefont {R.}~\bibnamefont {Attada}},\ }\bibfield  {title} {\enquote {\bibinfo {title} {Regional and temporal variability of indian summer monsoon rainfall in relation to el ni{\~n}o southern oscillation},}\ }\href {http://dx.doi.org/doi:10.1038/s41598-023-38730-5} {\bibfield  {journal} {\bibinfo  {journal} {Sci. Rep.}\ }\textbf {\bibinfo {volume} {13}},\ \bibinfo {pages} {12643} (\bibinfo {year} {2023})}\BibitemShut {NoStop}%
\bibitem [{\citenamefont {Sahana}\ \emph {et~al.}(2015{\natexlab{b}})\citenamefont {Sahana}, \citenamefont {Ghosh}, \citenamefont {Ganguly},\ and\ \citenamefont {Murtugudde}}]{47}%
  \BibitemOpen
  \bibfield  {author} {\bibinfo {author} {\bibfnamefont {A.}~\bibnamefont {Sahana}}, \bibinfo {author} {\bibfnamefont {S.}~\bibnamefont {Ghosh}}, \bibinfo {author} {\bibfnamefont {A.}~\bibnamefont {Ganguly}}, \ and\ \bibinfo {author} {\bibfnamefont {R.}~\bibnamefont {Murtugudde}},\ }\bibfield  {title} {\enquote {\bibinfo {title} {Shift in indian summer monsoon onset during 1976/1977},}\ }\href {http://dx.doi.org/doi:10.1088/1748-9326/10/5/054006} {\bibfield  {journal} {\bibinfo  {journal} {Environ. Res. Lett.}\ }\textbf {\bibinfo {volume} {10}},\ \bibinfo {pages} {054006} (\bibinfo {year} {2015}{\natexlab{b}})}\BibitemShut {NoStop}%
\bibitem [{\citenamefont {Xiang}\ and\ \citenamefont {Wang}(2013)}]{48}%
  \BibitemOpen
  \bibfield  {author} {\bibinfo {author} {\bibfnamefont {B.}~\bibnamefont {Xiang}}\ and\ \bibinfo {author} {\bibfnamefont {B.}~\bibnamefont {Wang}},\ }\bibfield  {title} {\enquote {\bibinfo {title} {Mechanisms for the advanced asian summer monsoon onset since the mid-to-late 1990s},}\ }\href {http://dx.doi.org/doi:10.1175/JCLI-D-12-00445.1} {\bibfield  {journal} {\bibinfo  {journal} {J. Clim.}\ }\textbf {\bibinfo {volume} {26}},\ \bibinfo {pages} {1993--2009} (\bibinfo {year} {2013})}\BibitemShut {NoStop}%
\bibitem [{\citenamefont {Ratna}\ \emph {et~al.}(2016)\citenamefont {Ratna}, \citenamefont {Cherchi}, \citenamefont {Joseph}, \citenamefont {Behera}, \citenamefont {Abish},\ and\ \citenamefont {Masina}}]{49}%
  \BibitemOpen
  \bibfield  {author} {\bibinfo {author} {\bibfnamefont {S.~B.}\ \bibnamefont {Ratna}}, \bibinfo {author} {\bibfnamefont {A.}~\bibnamefont {Cherchi}}, \bibinfo {author} {\bibfnamefont {P.}~\bibnamefont {Joseph}}, \bibinfo {author} {\bibfnamefont {S.}~\bibnamefont {Behera}}, \bibinfo {author} {\bibfnamefont {B.}~\bibnamefont {Abish}}, \ and\ \bibinfo {author} {\bibfnamefont {S.}~\bibnamefont {Masina}},\ }\bibfield  {title} {\enquote {\bibinfo {title} {Moisture variability over the indo-pacific region and its influence on the indian summer monsoon rainfall},}\ }\href {http://dx.doi.org/doi:10.1007/s00382-015-2624-z} {\bibfield  {journal} {\bibinfo  {journal} {Clim. Dyn.}\ }\textbf {\bibinfo {volume} {46}},\ \bibinfo {pages} {949--965} (\bibinfo {year} {2016})}\BibitemShut {NoStop}%
\bibitem [{\citenamefont {Polanski}\ \emph {et~al.}(2014)\citenamefont {Polanski}, \citenamefont {Fallah}, \citenamefont {Befort}, \citenamefont {Prasad},\ and\ \citenamefont {Cubasch}}]{50}%
  \BibitemOpen
  \bibfield  {author} {\bibinfo {author} {\bibfnamefont {S.}~\bibnamefont {Polanski}}, \bibinfo {author} {\bibfnamefont {B.}~\bibnamefont {Fallah}}, \bibinfo {author} {\bibfnamefont {D.~J.}\ \bibnamefont {Befort}}, \bibinfo {author} {\bibfnamefont {S.}~\bibnamefont {Prasad}}, \ and\ \bibinfo {author} {\bibfnamefont {U.}~\bibnamefont {Cubasch}},\ }\bibfield  {title} {\enquote {\bibinfo {title} {Regional moisture change over india during the past millennium: A comparison of multi-proxy reconstructions and climate model simulations},}\ }\href {http://dx.doi.org/doi:10.1016/j.gloplacha.2014.08.016} {\bibfield  {journal} {\bibinfo  {journal} {Glob. Planet. Change}\ }\textbf {\bibinfo {volume} {122}},\ \bibinfo {pages} {176--185} (\bibinfo {year} {2014})}\BibitemShut {NoStop}%
\bibitem [{\citenamefont {Gushchina}\ \emph {et~al.}(2020)\citenamefont {Gushchina}, \citenamefont {Zheleznova}, \citenamefont {Osipov},\ and\ \citenamefont {Olchev}}]{51}%
  \BibitemOpen
  \bibfield  {author} {\bibinfo {author} {\bibfnamefont {D.}~\bibnamefont {Gushchina}}, \bibinfo {author} {\bibfnamefont {I.}~\bibnamefont {Zheleznova}}, \bibinfo {author} {\bibfnamefont {A.}~\bibnamefont {Osipov}}, \ and\ \bibinfo {author} {\bibfnamefont {A.}~\bibnamefont {Olchev}},\ }\bibfield  {title} {\enquote {\bibinfo {title} {Effect of various types of enso events on moisture conditions in the humid and subhumid tropics},}\ }\href {http://dx.doi.org/doi:10.3390/atmos11121354} {\bibfield  {journal} {\bibinfo  {journal} {Atmosphere}\ }\textbf {\bibinfo {volume} {11}},\ \bibinfo {pages} {1354} (\bibinfo {year} {2020})}\BibitemShut {NoStop}%
\bibitem [{\citenamefont {Donges}\ \emph {et~al.}(2015)\citenamefont {Donges}, \citenamefont {Heitzig}, \citenamefont {Beronov}, \citenamefont {Wiedermann}, \citenamefont {Runge}, \citenamefont {Feng}, \citenamefont {Tupikina}, \citenamefont {Stolbova}, \citenamefont {Donner}, \citenamefont {Marwan}, \citenamefont {Dijkstra},\ and\ \citenamefont {Kurths}}]{pyuni}%
  \BibitemOpen
  \bibfield  {author} {\bibinfo {author} {\bibfnamefont {J.~F.}\ \bibnamefont {Donges}}, \bibinfo {author} {\bibfnamefont {J.}~\bibnamefont {Heitzig}}, \bibinfo {author} {\bibfnamefont {B.}~\bibnamefont {Beronov}}, \bibinfo {author} {\bibfnamefont {M.}~\bibnamefont {Wiedermann}}, \bibinfo {author} {\bibfnamefont {J.}~\bibnamefont {Runge}}, \bibinfo {author} {\bibfnamefont {Q.-Y.}\ \bibnamefont {Feng}}, \bibinfo {author} {\bibfnamefont {L.}~\bibnamefont {Tupikina}}, \bibinfo {author} {\bibfnamefont {V.}~\bibnamefont {Stolbova}}, \bibinfo {author} {\bibfnamefont {R.~V.}\ \bibnamefont {Donner}}, \bibinfo {author} {\bibfnamefont {N.}~\bibnamefont {Marwan}}, \bibinfo {author} {\bibfnamefont {H.~A.}\ \bibnamefont {Dijkstra}}, \ and\ \bibinfo {author} {\bibfnamefont {J.}~\bibnamefont {Kurths}},\ }\bibfield  {title} {\enquote {\bibinfo {title} {Unified functional network and nonlinear time series analysis for complex systems science: The pyunicorn package},}\ }\href {\doibase 10.1063/1.4934554} {\bibfield  {journal}
  {\bibinfo  {journal} {Chaos}\ }\textbf {\bibinfo {volume} {25}},\ \bibinfo {pages} {113101} (\bibinfo {year} {2015})}\BibitemShut {NoStop}%
\end{thebibliography}%

\clearpage
\newpage


\section*{Supplementary material}

# Supplementary material
# Recurrence analysis of meteorological data from climate zones in India

Joshin John Bejoy[1] and G. Ambika*[2]

[1]Indian Institute of Science Education and Research (IISER) Tirupati, Tirupati-517507 India
[1]Department of Aerospace Engineering, Indian Institute of Technology Madras, Chennai 600036, India
[2]Indian Institute of Science Education and Research (IISERTVM), Thiruvananthapuram-695551 India, *Corresponding Author

February 21, 2024

## 1 Variations in Recurrence measures

In this section we present details of variations in the RQA measures DET and LAM in each of the 15 locations. The values computed using the sliding window approach over the span of data as explained in the main text are shown in Fig:1,2 for temperature data and Fig:3,4 for relative humidity data.

The measures DET and LAM are estimated from the frequency distributions of diagonal and horizontal lines $P(l)$ and $P(v)$. The minimum lengths $l_{min}$ and $v_{min}$ used are those corresponding to the first maximum of $P(l) \times l$ and $P(v) \times v$ respectively.

The presented values are averages from two consecutive windows, and the time points shown correspond to their midpoints. The selected quantiles are indicated as dotted horizontal lines of so that significant trends can be identified. The variations outside the inter quantile range [0.1-0.9], are considered significant.

The variations in the recurrence network measures CPL and LD in each of the 15 locations are given in Fig:5 for temperature data and Fig:6 for relative humidity data. They are grouped into five sets based on the clusters obtained from cross-correlation analysis of variations in the measures over the period of the study.

# 2 Variations in RQA Measures

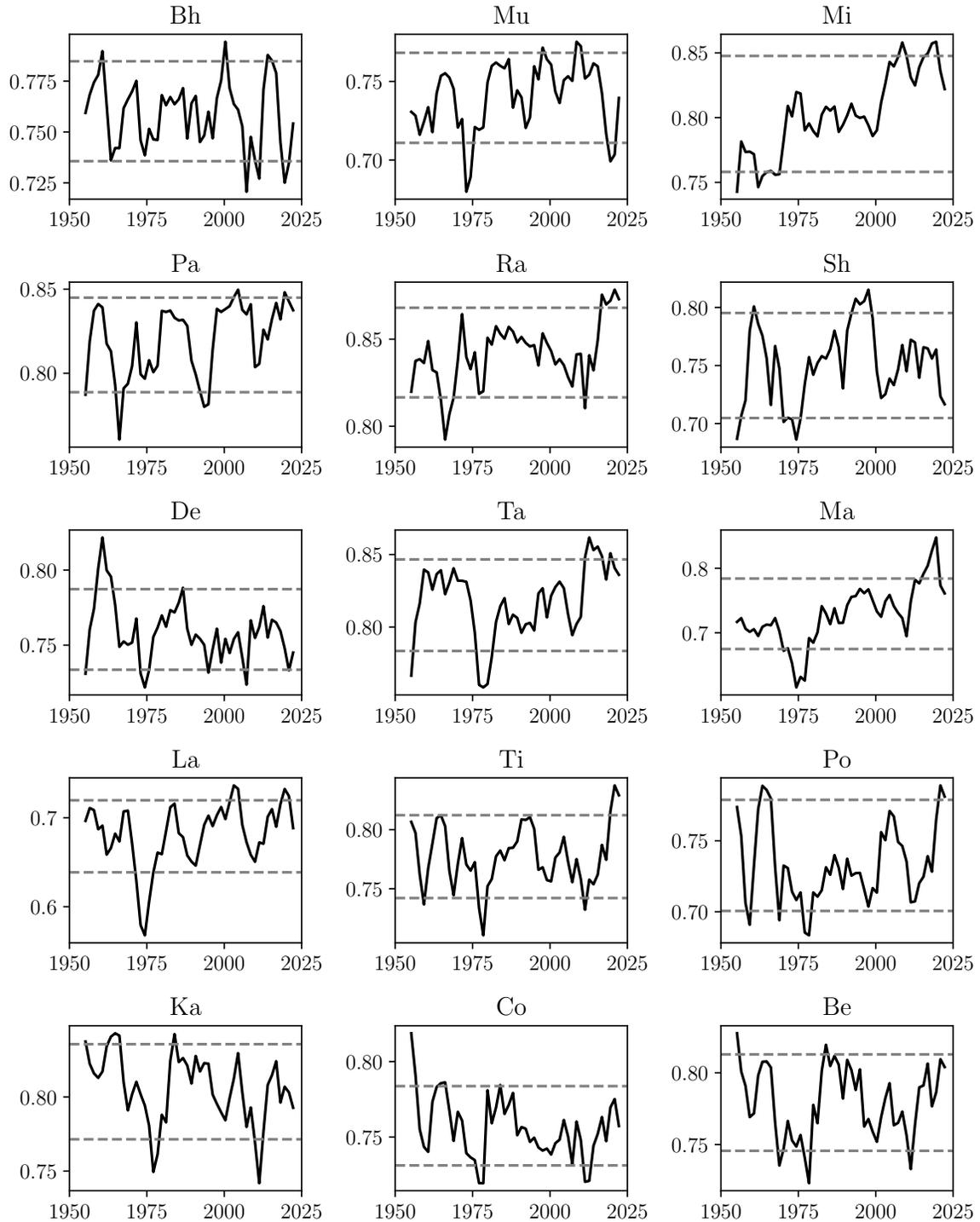

Figure 1: Variations in the recurrence measures DET($l_{min} = 15$ in each of the 15 locations from temperature data. The selected quantiles in the range [0.1,0.9] are indicated as the region between dotted lines, and the variations outside that are considered significant.

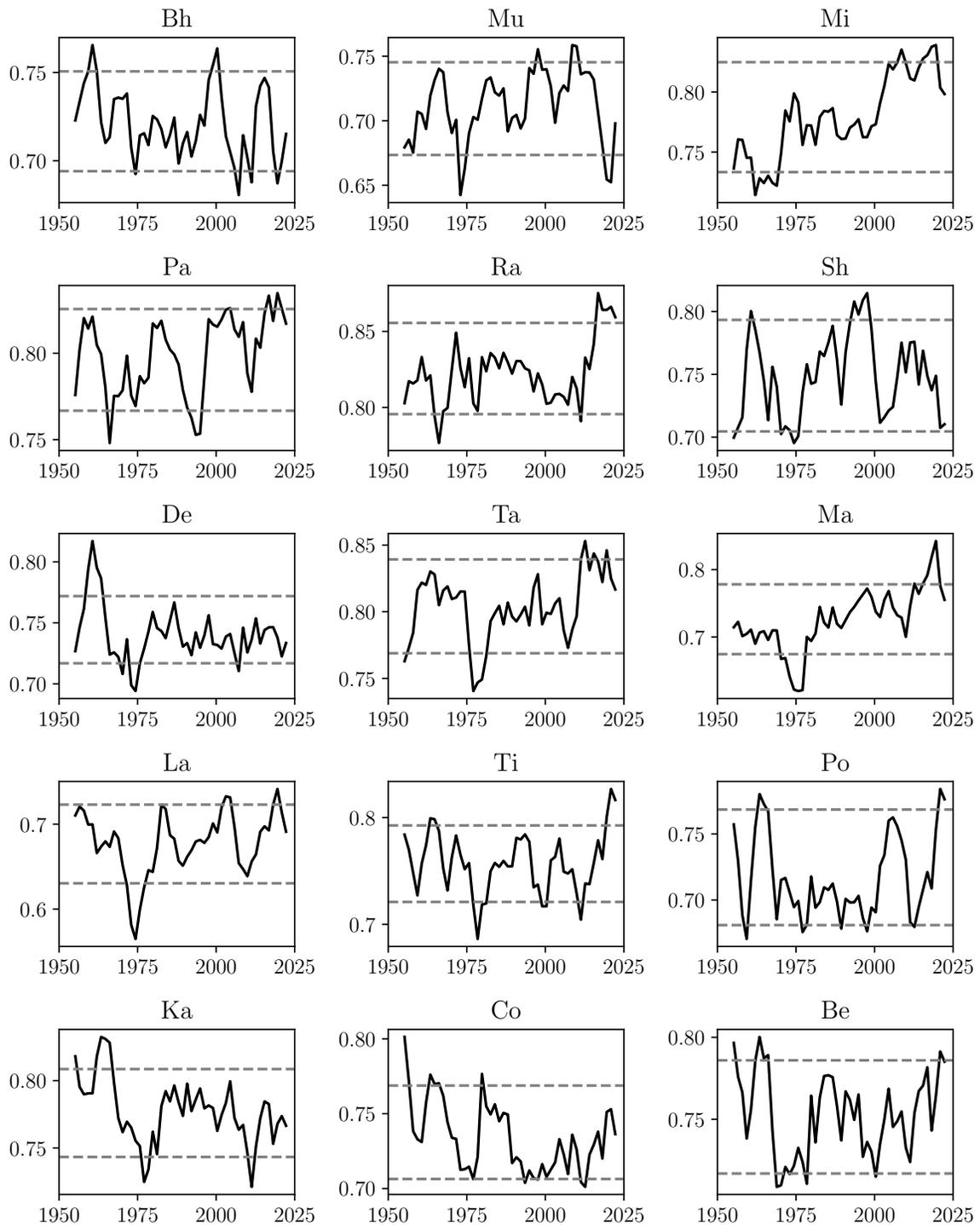

Figure 2: Variations in the recurrence measures LAM($v_{min} = 23$) in each of the 15 locations from temperature data. The selected quantiles in the range [0.1,0.9] are indicated as the region between dotted lines and the variations outside that are considered significant.

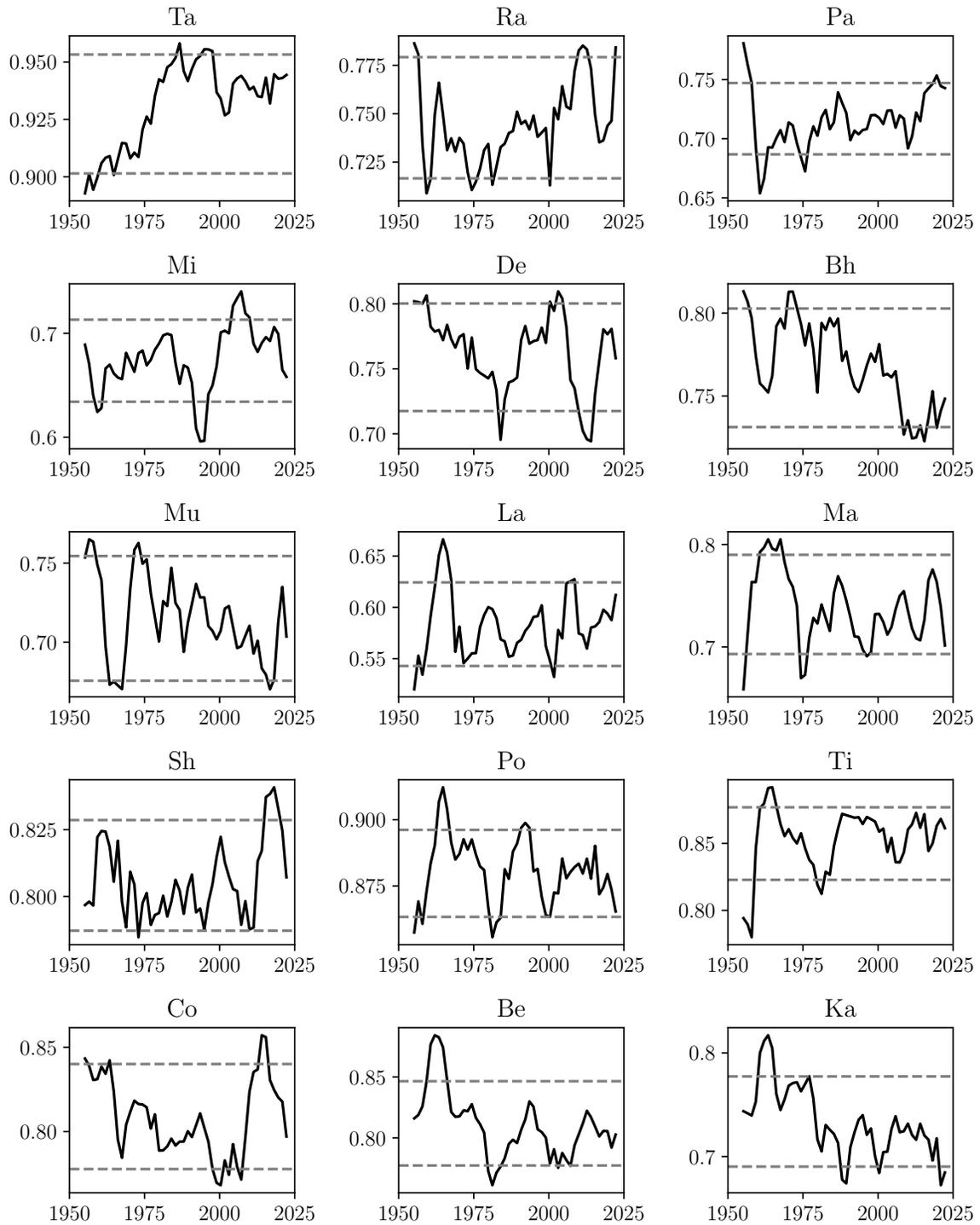

Figure 3: Variations in the recurrence measures DET($l_{min} = 10$) in each of the 15 locations from relative humidity data. The dotted lines indicate the selected quantiles [0.1-0.9].

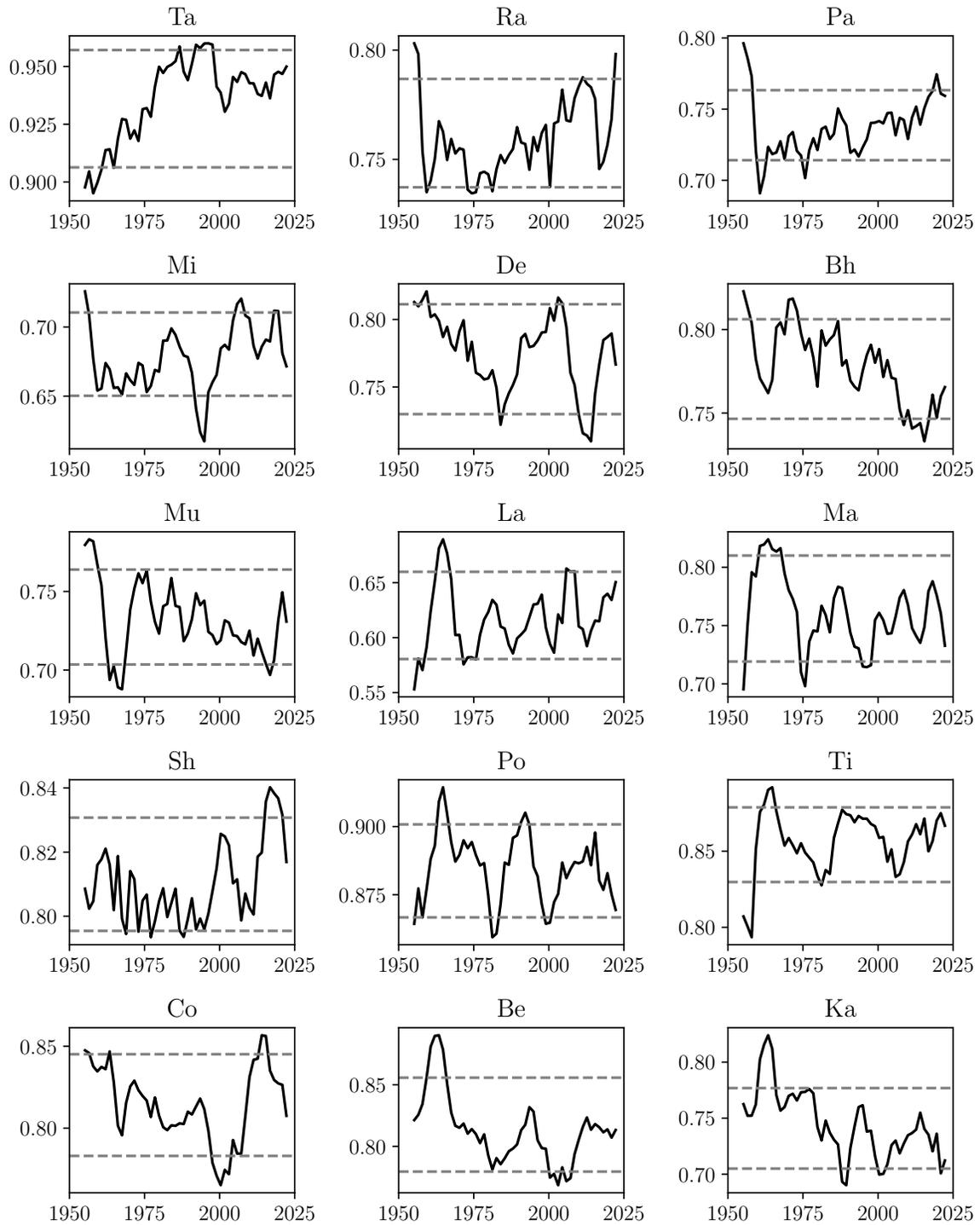

Figure 4: Variations in the recurrence measures LAM($v_{min} = 14$) in each of the 15 locations from relative humidity data. The dotted lines indicate the selected quantiles [0.1-0.9].

# 3 Variations in recurrence network measures

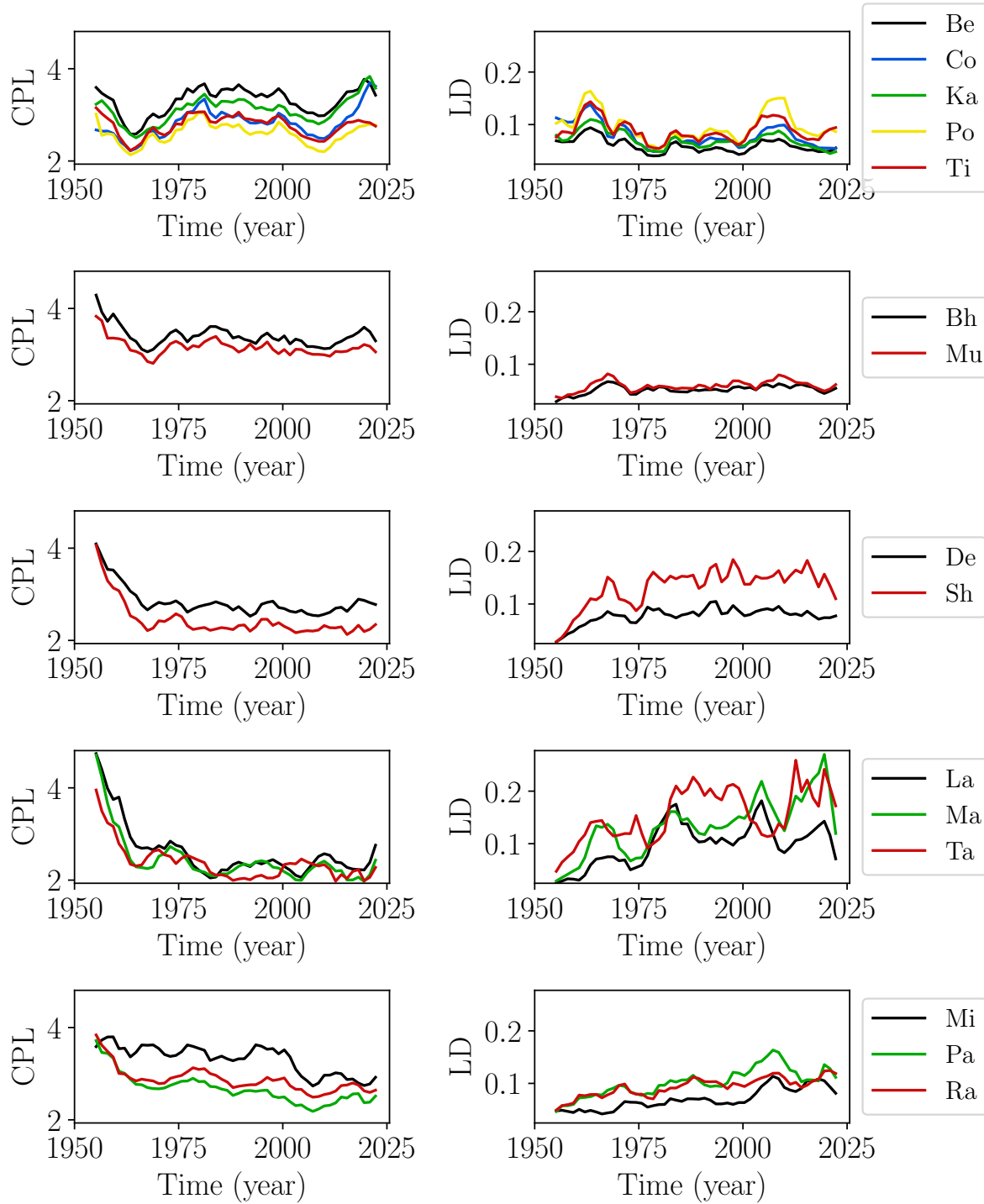

Figure 5: Variations in the recurrence network measure (a) CPL and (b) LD from temperature data for the five groups of locations.

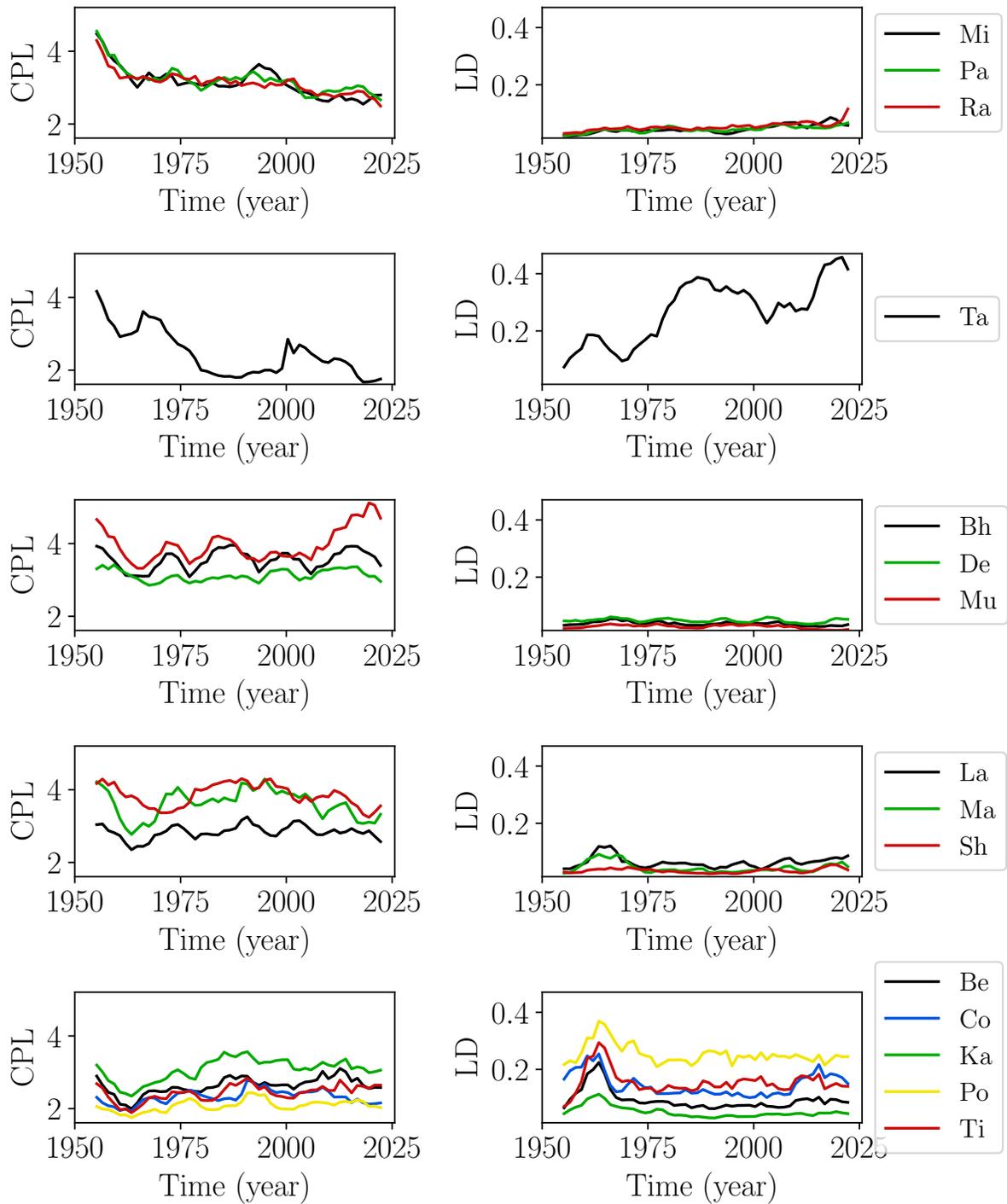

Figure 6: Variations in the recurrence network measure (a) CPL and (b) LD from relative humidity data for the five groups of locations.



\end{document}